\documentclass[
a4paper,reqno]{amsart}
\usepackage[T1]{fontenc}
\usepackage[english]{babel}
\usepackage{amssymb,bbm,enumerate}
\usepackage[normalem]{ulem}
\usepackage{color}
\usepackage[linktocpage=true,colorlinks=true, linkcolor=blue, citecolor=red, urlcolor=blue]{hyperref}
\usepackage[all]{xy}
\usepackage{graphicx}

\renewcommand{\phi}{\varphi}

\newcommand{\mc}[1]{\mathcal{#1}}
\newcommand{\mf}[1]{\mathfrak{#1}}
\newcommand{\mb}[1]{\mathbb{#1}}

\newcommand{\id}{\mathbbm{1}}
\newcommand{\quot}[2] {\ensuremath{\raisebox{.40ex}{\ensuremath{#1}}
\! \big / \! \raisebox{-.40ex}{\ensuremath{#2}}}}
\newcommand{\tint}{{\textstyle\int}}
\newcommand{\cmd}[1]{\texttt{#1}}
\newcommand{\dev}{\partial}

\DeclareMathOperator{\Mat}{Mat}

\DeclareMathOperator{\tr}{Tr}

\DeclareMathOperator{\ad}{ad}

\DeclareMathOperator{\rank}{rank}

\theoremstyle{plain}
\newtheorem{theorem}{Theorem}[section]

\newtheorem{proposition}[theorem]{Proposition}
\newtheorem{corollary}[theorem]{Corollary}

\theoremstyle{definition}
\newtheorem{definition}[theorem]{Definition}
\newtheorem{example}[theorem]{Example}

\theoremstyle{remark}
\newtheorem{remark}[theorem]{Remark}

\numberwithin{equation}{section}

\definecolor{light}{gray}{.9}

\title{M\lowercase{aster}PVA and WA\lowercase{lg}: Mathematica packages for Poisson vertex algebras and classical affine $\mc W$-algebras}
\author{Matteo Casati}
\address{Department of Mathematical Sciences, Loughborough University, Loughborough LE11 3TU, United Kingdom\newline Marie Curie fellow of the Istituto Nazionale d'Alta Matematica, 00185 Roma, Italy
}
\email{M.Casati@lboro.ac.uk}
\author{Daniele Valeri}
\address{
Yau Mathematical Sciences Center, Tsinghua University, 100084 Beijing, China
}
\email{daniele@math.tsinghua.edu.cn}

\begin{document}

\pagestyle{plain}

\begin{abstract}
We give an introduction to the Mathematica packages \textsl{MasterPVA} and \textsl{MasterPVAmulti}
used to compute $\lambda$-brackets in Poisson vertex algebras, which play an important role in the theory
of infinite-dimensional Hamiltonian systems.
As an application, we give an introduction to the Mathematica package \textsl{WAlg} aimed to compute the 
$\lambda$-brackets among the generators of classical affine $\mc W$-algebras.
The use of these packages is shown by providing some explicit examples.
\end{abstract}

\maketitle

\tableofcontents

\section{Introduction}

Poisson vertex algebras (PVA) arise as the quasi-classical limit of a family of vertex algebras \cite{DSK06} in the same way as Poisson
algebras arise as the quasi-classical limit of a family of associative algebras. 

Note also that  a PVA is a local counterpart of a Coisson (=chiral Poisson) algebra defined in \cite{BD04}.
Moreover, a PVA can be obtained as a formal Fourier transform of a local Poisson bracket \cite{BDSK09}, which
plays an important role in the theory of infinite-dimensional integrable Hamiltonian systems.
In fact, as demonstrated in \cite{BDSK09}, the language of $\lambda$-brackets \cite{D'AK98,Kac98} in the framework of
Poisson vertex algebras is often more convenient and transparent than the equivalent languages of local Poisson brackets,
used in the book \cite{FT86}, or of Hamiltonian operators, used in the book \cite{Dor93} (and references therein).

Hence, the theory of PVA has been extensively used in order to get a better understanding of generalized Drinfeld-Sokolov
hierarchies for classical affine $\mc W$-algebras \cite{DSKV13,DSKV14,DSKVnew}, Adler-Gelfand-Dickey hierarchies \cite{DSKV15} 
and, more generally, Lax type integrable Hamiltonian equations \cite{DSKVold},
and the Lenard-Magri scheme of integrability \cite{DSKT14,DSKT15}.
Furthermore, the notion of a PVA has been extended in \cite{CasPhD,Cas15} to deal with Hamiltonian operators, or, equivalently, local Poisson
brackets, for multidimensional systems of PDEs (namely, PDEs for functions depending on several spatial variables). The notion of multidimensional PVA has been used for studying the theory of symmetries and deformations of the so-called Poisson brackets of hydrodynamic type \cite{DN83}, as well as for the local nonlinear brackets associated with 2D Euler's equation \cite{Cas15b}.

\smallskip

One of the most remarkable accomplishments of the theory of PVA has been the
derivation
of an explicit formula for the bi-Hamiltonian structure underlying classical $\mc W$-algebras
\cite{DSKV16}.
Classical affine $\mc W$-algebras are associated to a pair $(\mf g,f)$ consisting of a
simple Lie algebra $\mf g$ and a nilpotent element $f\in\mf g$.
For a principal nilpotent element $f\in\mf g$, they appeared in the seminal paper by
Drinfeld and Sokolov \cite{DS85}. They were introduced as Poisson algebras of functions
on an infinite dimensional Poisson manifold,
and they were used to study KdV-type integrable
bi-Hamiltonian hierarchies of PDE's, nowadays known as Drinfeld-Sokolov hierarchies. 
Subsequently, in the 90's, there was an extensive literature extending the Drinfeld-Sokolov 
construction of classical $\mc W$-algebras and the corresponding generalized Drinfeld-Sokolov 
hierarchies to other nilpotent elements, \cite{dGHM92,FHM92,BdGHM93,DF95,FGMS95, FGMS96}.
Recently \cite{DSKV13}, classical affine $\mc W$-algebras were described as PVAs.
The powerful tool of the language of $\lambda$-bracket has been then used to get an explicit
formula for the bi-Hamiltonian structure describing them and to give a rigorous
definition, and to compute explicitly, generalized Drinfeld-Sokolov hierarchies \cite{DSKV14,DSKVnew}.
These results may find interesting applications in studying the relations of Drinfeld-Sokolov
hierarchies with Kac-Wakimoto hierarchies \cite{KW89} and
computation of the corresponding tau-functions,
and in the problem of quantization of classical integrable systems \cite{BLZ96} and applications to CFT.   

\smallskip

The most powerful tool in the PVA theory is the so-called Master Formula \eqref{masterformula}. It allows  to rephrase
the relevant questions in the theory of infinite-dimensional Hamiltonian systems in terms of the $\lambda$-bracket language,
thus providing a completely algebraic computational technique, which replaces all the manipulations used in \cite{FT86,Dor93}
in the setting of the formal calculus of variations.

Note that, in the $\lambda$-bracket language the computations are not necessarily hard to perform by hand,
but their numbers increase dramatically with the growing number of spatial dimensions.

The package \textsl{MasterPVA} and its generalization to the multidimensional case \textsl{MasterPVAmulti} have been written
to exploit a Computer Algebra System, like Mathematica, to automatically compute the Master Formula for PVA.
The choice of Mathematica is motivated by the pre-existing package \textsl{Lambda}, by J.~Ekstrand \cite{Eks11},
aimed to compute operator product expansions in conformal field theory using the $\lambda$-bracket language
within the framework of  vertex algebras \cite{Kac98,DSK06}.

These packages have been used in \cite{CasPhD} -- with some preliminary results published in \cite{Cas15} -- in order to compute up to second dispersive order the Casimir functions, the symmetries and the compatible deformations of the bidimensional Poisson brackets of hydrodynamic type. They have been proved effective also when working with scalar structures, whenever explicit computations are needed \cite{CCS15}.

The Mathematica package \textsl{WAlg} provides the implementation of the results about the structure theory of classical affine $\mc W$-algebras obtained in \cite{DSKV16}. It can be used
to compute all
$\lambda$-brackets between generators of the classical affine $\mc W$-algebras $\mc W(\mf g,f)$, where $\mf{g}$ is a simple Lie algebra of type $A,B,C,D$ and $G$, and $f\in\mf g$ is an arbitrary nilpotent element.
Thus we can obtain explicit expressions for the generalized Drinfeld-Sokolov hierarchies and
their bi-Hamiltonian structure by combining the programs \textsl{WAlg} and \textsl{MasterPVA}.

\smallskip

The paper is organized as follows.

In Section \ref{sec:pva} we review the definition of PVA following \cite{BDSK09} and its multidimensional generalization given in \cite{Cas15}.
In particular we introduce the notion of an algebra of differential functions and the Master Formula \eqref{masterformula}
used to perform $\lambda$-brackets computations on it, and we show that a PVA is equivalent to the notion of an Hamiltonian
operator (differently from \cite{Dor93} we call this Hamiltonian operator a Poisson structure). We also recall the connection with infinite-dimensional
Hamiltonian systems.

In Section \ref{sec:dsred} we review the definition of classical affine $\mc W$-algebras using the language of PVA,
following \cite{DSKV13}.
The main results are Theorems \ref{thm:structure-W} and \ref{20140304:thm} which give an explicit description
of, respectively, the differential algebra structure and the Poisson
structure of classical affine $\mc W$-algebras, see also \cite{DSKV16}.

In Section \ref{sec:3} we explain how to use the packages \textsl{MasterPVA} and \textsl{MasterPVAmulti} by giving some
explicit examples.
We show the well-known compatibility between GFZ and Virasoro-Magri PVA,
we derive the Dubrovin-Novikov conditions for a bidimensional Poisson structure of hydrodynamic type
\cite{DN83}, and, finally, we reprove the Mokhov's classification for the $N=1$ multidimensional structures of hydrodynamic type  \cite{M88}.

In Section \ref{sec:WAlg} we explain how to use the package \emph{WAlg} by giving some explicit examples. First, we consider
the case of a principal nilpotent element $f$ in the Lie algebra $\mf g=\mf o_7$ and we show how to compute a basis $\{q_j\}_{j\in J}$ 
of $\mf g^f$ and the corresponding set of generators $\{w(q_j)\}_{j\in J}$ of the classical affine $\mc W$-algebra
$\mc W(\mf g,f)$ given by Theorem \ref{thm:structure-W}.
Then, we consider the case of a minimal nilpotent element in the Lie algebra $\mf{sp}_4$ and 
we show how to compute the $\lambda$-brackets among the generators of the corresponding classical affine $\mc W$-algebra 
using Theorem \ref{20140304:thm}.
Finally, we compute explicitly all classical affine $\mc W$-algebras $\mc W(\mf g,f)$
corresponding to a simple Lie algebra $\mf g$ of rank $2$ and its principal nilpotent element $f$
and we compare our results with the ones in \cite{DSKW10}.

The complete list of commands provided by the packages \textsl{MasterPVA} and\\ \textsl{MasterPVAmulti}
(respectively \textsl{WAlg})
is given in Section \ref{sec:4} (respectively Section \ref{sec:6}).

\subsubsection*{Acquiring the packages}
The packages have been developed with Mathematica 9.0 and can be downloaded from

\url{http://www.theatron.it/daniele/MasterPVA_files.tar.gz}

\noindent where it is also possible to find the related libraries and the examples provided in this paper.

\subsubsection*{Acknowledgments}
We wish to thank Alberto De Sole, Boris Dubrovin and Victor Kac for introducing us to the
fascinating theory of integrable systems.
Part of this work was done during the visit of the authors to the Department of Mathematics of
the University of Rome La Sapienza in January and February 2016, and to the Department of Mathematics and Applications of the University of Milan-Bicocca in January 2017. We wish to thank this institutions for the kind hospitality.
We also wish to thank theatrOn.it for hosting the packages files.

The first author is supported by the INdAM-Cofund-2012 grant ``MPoisCoho -- Poisson cohomology of multidimensional Hamiltonian structures'' .

The second author is supported by an NSFC 
``Research Fund for International Young Scientists'' grant.
\section{Poisson vertex algebras and Hamiltonian equations}\label{sec:pva}

In this section we review the connection between Poisson vertex algebras
and the theory of Hamiltonian equations as laid down in \cite{BDSK09}.

\subsection{Poisson vertex algebras}
Let $\mc V$ be a \emph{differential algebra}, namely a unital commutative associative algebra
over a field $\mb F$ of characteristic $0$, with a derivation $\partial:\mc V\to\mc V$.
\begin{definition}
\begin{enumerate}[(a)]
\item
A \emph{$\lambda$-bracket} on $\mc V$ is an $\mb F$-linear map
$\mc V\otimes\mc V\to\mb F[\lambda]\otimes\mc V$,
denoted by $f\otimes g\to\{f_\lambda g\}$,
satisfying \emph{sesquilinearity} ($f,g\in\mc V$):
\begin{equation}\label{sesqui}
\{\partial f_\lambda g\}=
-\lambda\{f_\lambda g\},\qquad 
\{f_\lambda\partial g\}=
(\lambda+\partial)\{f_\lambda g\}\,,
\end{equation} 
and the \emph{left and right Leibniz rules} ($f,g,h\in\mc V$):
\begin{align}\label{lleibniz}
\{f_\lambda gh\}&=
\{f_\lambda g\}h+\{f_\lambda h\}g,\\
\{fh_\lambda g\}&=
\{f_{\lambda+\partial}g\}_{\rightarrow}h+\{h_{\lambda+\partial}g\}_{\rightarrow}f\,,\label{rleibniz}
\end{align}
where we use the following notation: if
$\{f_\lambda g\}=\sum_{n\in\mb Z_+}\lambda^n c_n$,
then
$\{f_{\lambda+\partial}g\}_{\rightarrow}h=\sum_{n\in\mb Z_+}c_n(\lambda+\partial)^nh$.
\item
We say that the $\lambda$-bracket is \emph{skew-symmetric} if
\begin{equation}\label{skewsim}
\{g_\lambda f\}+\{f_{-\lambda-\partial}g\}=0\,,
\end{equation}
where, now, 
$\{f_{-\lambda-\partial}g\}=\sum_{n\in\mb Z_+}(-\lambda-\partial)^nc_n$
(if there is no arrow we move $\partial$ to the left).
\item
A \emph{Poisson vertex algebra} (PVA) is a differential algebra $\mc V$ endowed 
with a $\lambda$-bracket which is skew-symmetric and satisfies 
the following \emph{Jacobi identity} in $\mc V[\lambda,\mu]$ ($f,g,h\in\mc V$):
\begin{equation}\label{jacobi}
\{f_\lambda\{g_\mu h\}\}-
\{\{f_\lambda g\}_{\lambda+\mu}h\}-
\{g_\mu\{f_\lambda h\}\}=0\,.
\end{equation}
\end{enumerate}
\end{definition}
\begin{example}\label{affinePVA}
Let $\mf g$ be a Lie algebra over $\mb F$ with a symmetric invariant bilinear form $\kappa$, 
and let $s$ be an element of $\mf g$.
The \emph{affine PVA}  associated to the triple $(\mf g,\kappa,s)$,
is the algebra of differential polynomials $\mc V=S(\mb F[\partial]\mf g)$ 
(where $\mb F[\partial]\mf g$ is the free $\mb F[\partial]$-module generated by $\mf g$
and $S(R)$ denotes the symmetric algebra over the $\mb F$-vector space $R$)
together with the $\lambda$-bracket given by
\begin{equation}\label{affinebracket}
\{a_\lambda b\}=[a,b]+\kappa(s|[a,b])+\kappa(a| b)\lambda
\qquad\text{ for } a,b\in\mf g
\,,
\end{equation}
and extended to $\mc V$ by sesquilinearity and the left and right Leibniz rules.
\end{example}

\subsection{Poisson vertex algebra structures on algebras of differential functions}\label{sec:1.2}

The basic examples of differential algebras are
the \emph{algebras of differential polynomials} in the variables $u_1,\ldots,u_{\ell}$:
$$
\mc R_\ell=\mb F[u_i^{(n)}\mid i\in I=\{1,\ldots,\ell\},n\in\mb Z_+]\,,
$$
where $\partial$ is the derivation defined by $\partial(u_i^{(n)})=u_i^{(n+1)}$, $i\in I,n\in\mb Z_+$.
Note that we have in $\mc V$ the following commutation relations:
\begin{equation}\label{partialcomm}
\left[\frac{\partial}{\partial u_i^{(n)}},\partial\right]=\frac{\partial}{\partial u_i^{(n-1)}}\,,
\end{equation}
where the RHS is considered to be zero if $n=0$.

An \emph{algebra of differential functions} in the variables $u_1,\dots,u_\ell$ is a differential algebra extension $\mc V$ of $\mc R_\ell$, endowed
with commuting derivations
$$
\frac{\partial}{\partial u_{i}^{(n)}}:\mc V\to\mc V\,,\qquad i\in I\,,\, n\in\mb Z_+\,,
$$
extending the usual partial derivatives on $\mc R_\ell$, such that only a finite number
of $\frac{\partial f}{\partial u_{i}^{(n)}}$ are non-zero for each $f\in\mc V$, and such that the commutation relations
\eqref{partialcomm} hold on $\mc V$.

The \emph{variational derivative} of $f\in\mc V$ with respect to $u_i$ is, by definition,
$$
\frac{\delta f}{\delta u_i}=\sum_{n\in\mb Z_+}(-\partial)^n\frac{\partial f}{\partial u_i^{(n)}}\,.
$$
The following result explains how to extend an arbitrary $\lambda$-bracket on a set of variables $\{u_i\}_{i\in I}$,
with value in some algebra of differential functions $\mc V$, to a PVA structure on $\mc V$.
\begin{theorem}[{\cite[Theorem 1.15]{BDSK09}}]\label{master}
Let $\mc V$ be an algebra of differential functions in the variables $\{u_i\}_{i\in I}$,
and let $H_{ij}(\lambda)\in\mb F[\lambda]\otimes\mc V,\,i,j\in I$.
\begin{enumerate}[(a)]
\item 
The Master Formula
\begin{equation}\label{masterformula}
\{f_\lambda g\}=
\sum_{\substack{i,j\in I\\m,n\in\mb Z_+}}\frac{\partial g}{\partial u_j^{(n)}}(\lambda+\partial)^n
H_{ji}(\lambda+\partial)(-\lambda-\partial)^m\frac{\partial f}{\partial u_i^{(m)}}
\end{equation}
defines a $\lambda$-bracket on $\mc V$
with given $\{u_i{}_\lambda u_j\}=H_{ji}(\lambda),\,i,j\in I$.
\item 
The $\lambda$-bracket \eqref{masterformula} on $\mc V$ satisfies the skew-symmetry 
condition \eqref{skewsim} provided that the same holds on generators ($i,j\in I$):
\begin{equation}\label{skewsimgen}
\{u_i{}_\lambda u_j\}+\{u_j{}_{-\lambda-\partial}u_i\}=0\,.
\end{equation}
\item 
Assuming that the skew-symmetry condition \eqref{skewsimgen} holds, 
the $\lambda$-bracket \eqref{masterformula} satisfies the Jacobi identity \eqref{jacobi}, 
thus making $\mc V$ a PVA, provided that the Jacobi identity holds 
on any triple of generators ($i,j,k\in I$):
\begin{equation}\label{jacobigen}
\{u_i{}_\lambda\{u_j{}_\mu u_k\}\}
-\{u_j{}_\mu\{u_i{}_\lambda u_k\}\}
-\{\{u_i{}_\lambda u_j\}_{\lambda+\mu}u_k\}
=0\,.
\end{equation}
\end{enumerate}
\end{theorem}
By Theorem \ref{master}(a), if $\mc V$ is an algebra of differential functions
in the variables $\{u_i\}_{i\in I}$, there is a bijective correspondence between
$\ell\times\ell$-matrices 
$H(\lambda)=
\left(H_{ij}(\lambda)\right)_{i,j\in I}\in\Mat_{\ell\times \ell}\mc V[\lambda]$
and the $\lambda$-brackets $\{\cdot\,_\lambda\,\cdot\}_H$ on $\mc V$ defined by the Master formula \eqref{masterformula}.
\begin{definition}\label{hamop}
A \emph{Poisson structure} on $\mc V$ is a matrix $H\in\Mat_{\ell\times\ell}\mc V[\lambda]$
such that the corresponding $\lambda$-bracket $\{\cdot\,_\lambda\,\cdot\}_H$
defines a PVA structure on $\mc V$.
\end{definition}

\begin{example}\label{affineHAMOP}
Consider the affine PVA defined in Example \ref{affinePVA}. 
Let $\{u_i\}_{i\in I}$ be a basis of $\mf g$.
The corresponding Poisson structure 
$H=\left(H_{ij}(\lambda)\right)\in\Mat_{\ell\times\ell}\mc V[\lambda]$ 
to the $\lambda$-bracket defined in \eqref{affinebracket} is given by
$$
H_{ij}(\lambda)=
\{u_j{}_{\lambda}u_i\}=
[u_j,u_i]+\kappa(s|[u_j,u_i])+\kappa(u_i| u_j)\lambda\,.
$$
\end{example}

\subsection{Poisson structures and Hamiltonian equations}

The relation between PVAs and Hamiltonian equations 
associated to a Poisson structure is based on the following simple observation.
\begin{proposition}\label{pvahamop}
Let $\mc V$ be a PVA. The $0$-th product on $\mc V$
induces a well defined Lie algebra bracket on the quotient space $\quot{\mc V}{\partial\mc V}$:
\begin{equation}\label{lambda=0}
\{\tint f,\tint g\}=\tint \left.\{f_\lambda g\}\right|_{\lambda=0}\,,
\end{equation}
where $\tint:\mc V\to\quot{\mc V}{\partial\mc V}$ is the canonical quotient map.
Moreover, we have a well defined Lie algebra action of $\quot{\mc V}{\partial\mc V}$ on $\mc V$
by derivations of the commutative associative product on $\mc V$, commuting with $\partial$,
given by 
$$
\{\tint f,g\}=\{f_\lambda g\}|_{\lambda=0}\,.
$$
\end{proposition}
In the special case when $\mc V$ is an algebra of differential functions in $\ell$ variables $\{u_i\}_{i\in I}$
and the PVA $\lambda$-bracket on $\mc V$ is associated to the Poisson structure 
$H\in\Mat_{\ell\times\ell}\mc V[\lambda]$,
the Lie bracket \eqref{lambda=0} on $\quot{\mc V}{\partial\mc V}$ takes the form 
(cf. \eqref{masterformula}):
\begin{equation}\label{liebrak}
\{\tint f,\tint g\}
=\sum_{i,j\in I}\int\frac{\delta g}{\delta u_j}H_{ji}(\partial)\frac{\delta f}{\delta u_i}
\,.
\end{equation}

\begin{definition}\label{hamsys}
Let $\mc V$ be an algebra of differential functions
with a Poisson structure $H$.
\begin{enumerate}[(a)]
\item 
Elements of $\quot{\mc V}{\partial\mc V}$ are called \emph{local functionals}. 
\item
Given a local functional $\int h\in\quot{\mc V}{\partial\mc V}$, 
the corresponding \emph{Hamiltonian equation} is
\begin{equation}\label{hameq}
\frac{du}{dt}
=\{\tint h,u\}_H
\qquad
\Big(\text{equivalently,}\quad
\frac{du_i}{dt}
=\sum_{j\in I}H_{ij}(\partial)\frac{\delta h}{\delta u_j},\ i\in I\Big)\,.
\end{equation}
\item 
A local functional $\int f\in\quot{\mc V}{\partial\mc V}$ is called an \emph{integral of motion} 
of equation \eqref{hameq} if $\frac{df}{dt}=0\mod\partial\mc V$
in virtue of \eqref{hameq}, or, equivalently, if $\tint h$ and $\tint f$ are \emph{in involution}:
$$
\{\tint h,\tint f\}_H=0\,.
$$
Namely, $\tint f$ lies in the centralizer of $\tint h$ in the Lie algebra $\quot{\mc V}{\partial\mc V}$
with Lie bracket \eqref{liebrak}.
\item 
Equation \eqref{hameq} is called \emph{integrable}
if there exists an infinite sequence $\tint f_0=\tint h,\,\tint f_1,\,\tint f_2,\dots$,
of linearly independent integrals of motion in involution.
The corresponding \emph{integrable hierarchy of Hamiltonian equations} is
\begin{equation}\label{eq:hierarchy}
\frac{du}{dt_n}=\{\tint f_n,u\}_H,\ n\in\mb Z_+
\,.
\end{equation}
(Equivalently,
$\frac{du_i}{dt_n}
=\sum_{j\in I}H_{ij}(\partial)\frac{\delta f_n}{\delta u_j}$, $n\in\mb Z_+,\,i\in I$)
\end{enumerate}
\end{definition}

\subsection{Multidimensional Poisson Vertex Algebras}\label{sec:multi}
The definition of a PVA has been extended in \cite{Cas15} in order to study Hamiltonian evolutionary PDEs with several spatial dimensions.

A \emph{$D$-dimensional differential algebra} is a unital commutative associative algebra $\mc V$
over a field $\mb F$ of characteristic $0$, endowed with $D$ commuting derivations $\partial_\alpha:\mc V\to\mc V$,
$\alpha=1\dots,D$.
\begin{definition}
\begin{enumerate}[(a)]
\item
A \emph{$D$-dimensional $\lambda$-bracket} on $\mc V$ is an $\mb F$-linear map
$\mc V\otimes\mc V\to\mb F[\lambda_1,\dots,\lambda_D]\otimes\mc V$,
denoted by $f\otimes g\to\{f_\lambda g\}$,
satisfying \emph{sesquilinearity} ($f,g\in\mc V$, $\alpha=1,\dots,D$):
\begin{equation}\label{sesqui_multi}
\{\partial_\alpha f_\lambda g\}=
-\lambda_\alpha\{f_\lambda g\},\qquad 
\{f_\lambda\partial_\alpha g\}=
(\lambda_\alpha+\partial_\alpha)\{f_\lambda g\}\,,
\end{equation} 
and the \emph{left and right Leibniz rules} \eqref{lleibniz} and \eqref{rleibniz}.
\item
We say that the $\lambda$-bracket is \emph{skew-symmetric} if
equation \eqref{skewsim} is satisifed.
\item
A \emph{$D$-dimensional PVA} is a $D$-dimensional differential algebra $\mc V$ endowed 
with a $\lambda$-bracket which is skew-symmetric and satisfies 
the \emph{Jacobi identity} \eqref{jacobi} in $\mc V[\lambda_1,\dots,\lambda_D,\mu_1,\dots,\mu_D]$.
\end{enumerate}
\end{definition}
The definition of an algebra of differential functions given in Section \ref{sec:1.2} can be generalized to the $D$-dimensional case 
and the analogous result to Theorem \ref{master} can be obtained (see \cite[Theorem 1]{Cas15}).

\section{Classical affine \texorpdfstring{$\mc W$}{W}-algebras}\label{sec:dsred}

In this section we recall the definition of classical affine $\mc W$-algebras $\mc W(\mf g,f)$ in the language 
of Poisson vertex algebras, following \cite{DSKV13}
(which is a development of \cite{DS85}).

\subsection{Setup and notation}\label{slod.4}

Let $\mf g$ be a simple Lie algebra with a non-degenerate symmetric invariant bilinear form $(\cdot\,|\,\cdot)$,
and let $\{f,2x,e\}\subset\mf g$ be an $\mf{sl}_2$-triple in $\mf g$.
We have the corresponding $\ad x$-eigenspace decomposition
$$
\mf g=\bigoplus_{k\in\frac{1}{2}\mb Z}\mf g_{k}
\,\,\text{ where }\,\,
\mf g_k=\big\{a\in\mf g\,\big|\,[x,a]=ka\big\}
\,.
$$
Clearly, $f\in\mf g_{-1}$, $x\in\mf g_{0}$ and $e\in\mf g_{1}$.
We let $d$ be the \emph{depth} of the grading, i.e. the maximal eigenvalue of $\ad x$.

By representation theory of $\mf{sl}_2$, the Lie algebra $\mf g$ admits the direct sum decompositions
\begin{equation}\label{20140221:eq4}
\mf g
=\mf g^f\oplus[e,\mf g]
=\mf g^e\oplus[f,\mf g]
\,.
\end{equation}
They are dual to each other, in the sense that $\mf g^f\perp[f,\mf g]$ and $[e,\mf g]\perp\mf g^e$.
For $a\in\mf g$, we denote by $a^\sharp=\pi_{\mf g^f}(a)\in\mf g^f$ its component in $\mf g^f$
with respect to the first decomposition in \eqref{20140221:eq4}.
Note that, since $[e,\mf g]$ is orthogonal to $\mf g^e$,
the spaces $\mf g^f$ and $\mf g^e$ are non-degenerately paired by $(\cdot\,|\,\cdot)$.

Next, we choose a basis of $\mf g$ as follows.
Let $\{q_j\}_{j\in J^f}$ be a basis of $\mf g^f$ consisting of $\ad x$-eigenvectors,
and let  $\{q^j\}_{j\in J^f}$ be the the dual basis of $\mf g^e$.
For $j\in J^f$,
we let $\delta(j)\in\frac12\mb Z$ be the $\ad x$-eigenvalue of $q^j$,
so that
\begin{equation}\label{20130520:eq5}
[x,q_j]=-\delta(j)q_j
\,\,,\,\,\,\,
[x,q^j]=\delta(j)q^j
\,.
\end{equation}
For $k\in\frac12\mb Z_+$
we also let $J^f_{-k}=\{i\in J^f\,|\,\delta(i)=k\}\subset J^f$,
so that $\{q_j\}_{j\in J^f_{-k}}$ is a basis of $\mf g^f_{-k}$,
and $\{q^j\}_{j\in J^f_{-k}}$ is the dual basis of $\mf g^e_{k}$.
By representation theory of $\mf{sl}_2$,
we get a basis of $\mf g$ consisting of the following elements:
\begin{equation}\label{20140221:eq1}
q^j_n=(\ad f)^nq^j
\,\,\text{ where }\,\,
n\in\{0,\dots,2\delta(j)\}
\,\,,\,\,\,\,
j\in J^f
\,.
\end{equation}
This basis consists of $\ad x$-eigenvectors,
and, for $k\in\frac12\mb Z$ such that $-d\leq k\leq d$,
the corresponding basis of $\mf g_k\subset\mf g$ is 
$\{q^j_{n}\}_{(j,n)\in J_{-k}}$,
where $J_{-k}$ is the following index set
\begin{equation}\label{20140221:eq5}
J_{-k}
=
\Big\{
(j,n)\in J^f\times\mb Z_+\,\Big|\,
\delta(j)-|k|\in\mb Z_+,\,n=\delta(j)-k
\Big\}
\,.
\end{equation}
The union of all these index sets is the index set for the basis of $\mf g$:
\begin{equation}\label{20140221:eq6}
J
=
\bigsqcup_{h\in\frac12\mb Z}J_h
=
\Big\{
(j,n)\,\Big|\,
j\in J^f,\,n\in\{0,\dots,2\delta(j)\}
\Big\}
\,.
\end{equation}

By \cite[Lemma 2.5]{DSKV16}, the corresponding basis of $\mf g$ dual to \eqref{20140221:eq1} is given
by ($(j,n)\in J$):
\begin{equation}\label{20140221:eq3}
q_j^n
=
\frac{(-1)^n}{(n!)^2\binom{2\delta(j)}{n}}
(\ad e)^nq_j
\,.
\end{equation}

Clearly, the bases \eqref{20140221:eq1} and \eqref{20140221:eq3} 
are compatible with the direct sum decompositions \eqref{20140221:eq4}.
In fact, we can write the corresponding projections
$\pi_{\mf g^f}$, $\pi_{[e,\mf g]}=1-\pi_{\mf g^f}$,
$\pi_{\mf g^e}$, and $\pi_{[f,\mf g]}=1-\pi_{\mf g^e}$,
in terms of these bases:
\begin{equation}\label{20130520:eq1}
\begin{array}{l}
\displaystyle{
\vphantom{Big(}
a^\sharp = \pi_{\mf g^f}(a)=\sum_{j\in J^f}(a|q^j)q_j
\,\,,\,\,\,\,
\pi_{[e,\mf g]}(a)=\sum_{j\in J^f}\sum_{n=1}^{2\delta(j)}(a|q^j_n)q_j^n
\,,} \\
\displaystyle{
\vphantom{Big(}
\pi_{\mf g^e}(a)=\sum_{j\in J^f}(a|q_j)q^j
\,\,,\,\,\,\,
\pi_{[f,\mf g]}(a)=\sum_{j\in J^f}\sum_{n=1}^{2\delta(j)}(a|q_j^n)q^j_n
\,.}
\end{array}
\end{equation}
Note that when $\delta(j)=0$, then the sums over $n$ in \eqref{20130520:eq1} become empty sums.

\subsection{Construction of the classical affine \texorpdfstring{$\mc W$}{W}-algebra}
\label{sec:2.1}

Recall from Example \ref{affinePVA} that
given an element $s\in\mf g$, we have a PVA structure on
the algebra of differential polynomials $\mc V(\mf g)=S(\mb F[\partial]\mf g)$,
with $\lambda$-bracket given on generators by 
\begin{equation}\label{lambda}
\{a_\lambda b\}_z=[a,b]+(a| b)\lambda+z(s|[a,b]),
\qquad a,b\in\mf g\,,
\end{equation}
and extended to $\mc V(\mf g)$ by the sesquilinearity axioms and the Leibniz rules.
Here $z$ is an element of the field $\mb F$.

We shall assume that $s$ lies in $\mf g_d$.
In this case the $\mb F[\partial]$-submodule
$\mb F[\partial]\mf g_{\geq\frac12}\subset\mc V(\mf g)$ 
is a Lie conformal subalgebra (see \cite{Kac98} for the definition)
with the $\lambda$-bracket $\{a_\lambda b\}_z=[a,b]$, $a,b\in\mf g_{\geq\frac12}$
(it is independent of $z$, since $s$ commutes with $\mf g_{\geq\frac12}$).
Consider the differential subalgebra
$\mc V(\mf g_{\leq\frac12})=S(\mb F[\partial]\mf g_{\leq\frac12})$ of $\mc V(\mf g)$,
and denote by $\rho:\,\mc V(\mf g)\twoheadrightarrow\mc V(\mf g_{\leq\frac12})$,
the differential algebra homomorphism defined on generators by
\begin{equation}\label{rho}
\rho(a)=\pi_{\leq\frac12}(a)+(f| a),
\qquad a\in\mf g\,,
\end{equation}
where $\pi_{\leq\frac12}:\,\mf g\to\mf g_{\leq\frac12}$ denotes 
the projection with kernel $\mf g_{\geq1}$.
Recall from \cite{DSKV13} that
we have a representation of the Lie conformal algebra $\mb F[\partial]\mf g_{\geq\frac12}$ 
on the differential subalgebra $\mc V(\mf g_{\leq\frac12})\subset\mc V(\mf g)$ given by
($a\in\mf g_{\geq\frac12}$, $g\in\mc V(\mf g_{\leq\frac12})$):
\begin{equation}\label{20120511:eq1}
a\,^\rho_\lambda\,g=\rho\{a_\lambda g\}_z
\end{equation}
(note that the RHS is independent of $z$ since, by assumption, $s\in Z(\mf g_{\geq\frac12})$).

The \emph{classical affine} $\mc W$-\emph{algebra} is, by definition,
the differential algebra
\begin{equation}\label{20120511:eq2}
\mc W=\mc W(\mf g,f)
=\big\{g\in\mc V(\mf g_{\leq\frac12})\,\big|\,a\,^\rho_\lambda\,g=0\,\text{ for all }a\in\mf g_{\geq\frac12}\}\,,
\end{equation}
endowed with the following PVA $\lambda$-bracket
\begin{equation}\label{20120511:eq3}
\{g_\lambda h\}_{z,\rho}=\rho\{g_\lambda h\}_z,
\qquad g,h\in\mc W\,.
\end{equation}
\begin{remark}\label{hierarchies_W}
Thinking of $z$ as a formal parameter, equation \eqref{20120511:eq3} gives a 1-parameter 
family of PVA structures on $\mc W$, or, equivalently, a bi-Poisson structure.
Indeed, we can write $\{g_\lambda h\}_{z,\rho}=\{g_\lambda h\}_{1,\rho}+z\{g_\lambda h\}_{0,\rho}$, for every $g,h\in\mc W$.
The $\lambda$-bracket $\{\cdot\,_\lambda\,\cdot\}_{1,\rho}$ does not depend on the choice
of $s\in\mf g_d$, while $\{\cdot\,_\lambda\,\cdot\}_{0,\rho}$ does.

Generalizing the results in \cite{DS85} it has been shown in \cite{DSKV13}, using the
Lenard-Magri scheme of integrability \cite{Mag78}, that it is possible to construct an integrable hierarchy of
bi-Hamiltonian equations for $\mc W$, known as generalized Drinfeld-Sokolov hierarchy, under the assumption that $f+s\in\mf g$ is a semisimple element.

Recently, generalized Drinfeld-Sokolov hierarchies for any nilpotent element $f\in\mf{gl}_N$ and
non-zero $s\in\mf g_d$ have been constructed in \cite{DSKVnew} using the theory of Adler type 
pseudodifferential operators \cite{DSKVold}.
\end{remark}

\subsection{Structure Theorem for classical affine \texorpdfstring{$\mc W$}{W}-algebras}
\label{sec:2.2}

In the algebra of differential polynomials $\mc V(\mf g_{\leq\frac12})$
we introduce the grading by \emph{conformal weight},
denoted by $\Delta\in\frac12\mb Z$, defined as follows.
For $a\in\mf g$ such that $[x,a]=\delta(a)a$, we let $\Delta(a)=1-\delta(a)$.
For a monomial $g=a_1^{(m_1)}\dots a_s^{(m_s)}$,
product of derivatives of $\ad x$ eigenvectors $a_i\in\mf g_{\leq\frac12}$,
we define its conformal weight as
\begin{equation}\label{degcw}
\Delta(g)=\Delta(a_1)+\dots+\Delta(a_s)+m_1+\dots+m_s\,.
\end{equation}
Thus we get the conformal weight space decomposition
$$
\mc V(\mf g_{\leq\frac12})=\bigoplus_{\Delta\in\frac12\mb Z_+}\mc V(\mf g_{\leq\frac12})\{\Delta\}\,.
$$
For example $\mc V(\mf g_{\leq\frac12})\{0\}=\mb F$,
$\mc V(\mf g_{\leq\frac12})\{\frac12\}=\mf g_{\frac12}$,
and $\mc V(\mf g_{\leq\frac12})\{1\}=\mf g_{0}\oplus S^2\mf g_{\frac12}$.

\begin{theorem}[\cite{DSKV13}]\label{daniele2}
Consider the PVA $\mc W=\mc W(\mf g,f)$ with the $\lambda$-bracket $\{\cdot\,_\lambda\,\cdot\}_{z,\rho}$
defined by equation \eqref{20120511:eq3}.
\begin{enumerate}[(a)]
\item
For every element $q\in\mf g^f_{1-\Delta}$ there exists a (not necessarily unique)
element $w\in\mc W\{\Delta\}=\mc W\cap\mc V(\mf g_{\leq\frac12})\{\Delta\}$ 
of the form $w=q+g$, where 
\begin{equation}\label{20140221:eq9}
g=\sum b_1^{(m_1)}\dots b_s^{(m_s)}\in\mc V(\mf g_{\leq\frac12})\{\Delta\}\,,
\end{equation}
is a sum of products of derivatives of $\ad x$-eigenvectors 
$b_i\in\mf g_{1-\Delta_i}\subset\mf g_{\leq\frac12}$,
such that
$$
\Delta_1+\dots+\Delta_s+m_1+\dots+m_s=\Delta
\,\,\text{ and }\,\,
s+m_1+\dots+m_s>1
\,.
$$
\item
Let $\{w_j=q_j+g_j\}_{j\in J^f}$ be any collection of elements in $\mc W$ as in part (a).
(Recall, from Section \ref{slod.4}, that $\{q_j\}_{j\in J^f}$ is a basis of $\mf g^f$ consisting of
$\ad x$-eigenvectors.)
Then the differential subalgebra $\mc W\subset\mc V(\mf g_{\leq\frac12})$ 
is the algebra of differential polynomials in the variables $\{w_j\}_{j\in J^f}$.
The algebra $\mc W$ is a graded associative algebra,
graded by the conformal weights defined in \eqref{degcw}:
$\mc W
=
\mb F\oplus\mc W\{1\}\oplus\mc W\{\frac32\}\oplus\mc W\{2\}\oplus\mc W\{\frac52\}\oplus\dots$.
\end{enumerate}
\end{theorem}

Recall the first of the direct sum decompositions \eqref{20140221:eq4}.
By assumption, the elements $q^0_j=q_j,\,j\in J^f$, form  a basis of $\mf g^f$,
and by construction the elements $q^n_j,\,(j,n)\in J$, with $n\geq1$,
form a basis of $[e,\mf g]$
(here we are using the notation from Section \ref{slod.4}).
Since $\mf g^f\subset\mf g_{\leq\frac12}$, we have the corresponding direct sum decomposition
\begin{equation}\label{20140221:eq7}
\mf g_{\leq\frac12}=\mf g^f\oplus[e,\mf g_{\leq-\frac12}]\,.
\end{equation}
It follows that the algebra of differential polynomials $\mc V(\mf g_{\leq\frac12})$
admits the following decomposition in a direct sum of subspaces
\begin{equation}\label{20140221:eq8}
\mc V(\mf g_{\leq\frac12})
=
\mc V(\mf g^f)
\oplus
\big\langle[e,\mf g_{\leq-\frac12}]\big\rangle_{\mc V(\mf g_{\leq\frac12})}
\,,
\end{equation}
where $\mc V(\mf g^f)$ is the algebra of differential polynomials over $\mf g^f$, 
and $\big\langle[e,\mf g_{\leq-\frac12}]\big\rangle_{\mc V(\mf g_{\leq\frac12})}$
is the differential ideal of $\mc V(\mf g_{\leq\frac12})$
generated by $[e,\mf g_{\leq-\frac12}]$.

Theorem \ref{daniele2} implies the following result.
\begin{corollary}[\cite{DSKV16}]\label{20140221:cor}
For every $q\in\mf g^f$ there exists a unique
element $w=w(q)\in\mc W$ 
of the form $w=q+r$, where 
$r\in\big\langle[e,\mf g_{\leq-\frac12}]\big\rangle_{\mc V(\mf g_{\leq\frac12})}$.
Moreover, if $q\in\mf g_{1-\Delta}$,
then $w(q)$ lies in $\mc W\{\Delta\}$ and $r$ is of the form \eqref{20140221:eq9}.
Consequently, $\mc W$ coincides with the algebra of differential polynomials in the variables
$w_j=w(q_j)$, $j\in J^f$.
\end{corollary}
As an immediate consequence of Theorem \ref{daniele2} and Corollary \ref{20140221:cor}
we get the following:
\begin{theorem}\label{thm:structure-W}
The map $\pi_{\mf g^f}$ restricts to a differential algebra isomorphism
$$
\pi:=\pi_{\mf g^f}|_{\mc W}:\,\mc W\,\stackrel{\sim}{\longrightarrow}\,\mc V(\mf g^f)
\,,
$$
hence we have the inverse differential algebra isomorphism
$$
w=:\,\mc V(\mf g^f)\,\stackrel{\sim}{\longrightarrow}\,\mc W
\,,
$$
which associates to every element $q\in\mf g^f$ the (unique) element $w(q)\in\mc W$
of the form $w(q)=q+r$, with $r\in\big\langle[e,\mf g_{\leq-\frac12}]\big\rangle_{\mc V(\mf g_{\leq\frac12})}$.
\end{theorem}

\subsection{Poisson structure of the classical affine $\mc W$-algebra}

Let $\ell=\dim\mf g^f$. By Corollary \ref{20140221:cor} the Poisson structure
$H=\left(H_{ij}(\lambda)\right)_{i,j\in J^f}\in\Mat_{\ell\times\ell}\mc W[\lambda]$ associated to the classical affine
$\mc W$-algebra $\mc W$
defined by equations \eqref{20120511:eq2} and \eqref{20120511:eq3} is given by ($i,j\in J^f$)
\begin{equation}\label{ham_W}
H_{ji}(\lambda)=\{ w(q_i)_\lambda w(q_j)\}_{z,\rho}\,.
\end{equation}

For $h,k\in\frac12\mb Z$, we introduce the notation
\begin{equation}\label{20140304:eq5}
h\prec k
\,\,\text{ if and only if }\,\,
h\leq k-1
\,.
\end{equation}
Also, for $t\geq1$, we denote $\vec{k}=(k_1,k_2,\dots,k_t)\in(\frac12\mb Z)^t$,
and $J_{-\vec{k}}:=J_{-k_1}\times\dots J_{-k_t}$.
Therefore, 
an element $(\vec{j},\vec{n})\in J_{-\vec{k}}$ is an $t$-tuple with
\begin{equation}\label{20140304:eq6}
(j_1,n_1)\in J_{-k_1},\dots,(j_t,n_t)\in J_{-k_t}
\,.
\end{equation}
The explicit expression of the Poisson structure $H$ defined by equation \eqref{ham_W} can be obtained by the following result.
\begin{theorem}[{\cite[Theorem 5.3]{DSKV16}}]\label{20140304:thm}
For $a\in \mf g^f_{-h}$ and $b\in\mf g^f_{-k}$, we have
\begin{equation}\label{20140304:eq4}
\begin{array}{l}
\displaystyle{
\phantom{\Big(}
\{w(a)_\lambda w(b)\}_{z,\rho}
=
w([a,b])+(a|b)\lambda+z(s|[a,b])
} \\
\displaystyle{
\phantom{\Big(}
-\sum_{t=1}^\infty
\sum_{
-h+1\leq k_t\prec \dots\prec k_1\leq k
}
\sum_{
(\vec{j},\vec{n})\in J_{-\vec{k}}
}
\big(
w([b,q^{j_1}_{n_1}]^\sharp)
-(b|q^{j_1}_{n_1})(\lambda+\partial)
+z(s|[b,q^{j_{1}}_{n_{1}}])
\big)
} \\
\displaystyle{
\phantom{\Big(}
\times\big(
w([q^{n_1+1}_{j_1},q^{j_2}_{n_2}]^\sharp)
-(q^{n_1+1}_{j_1}|q^{j_2}_{n_2}) (\lambda+\partial)
+z(s|[q^{n_1+1}_{j_1},q^{j_2}_{n_2}])
\big)
\dots } \\
\displaystyle{
\phantom{\Big(}
\dots
\times\big(
w([q^{n_{t-1}+1}_{j_{t-1}},q^{j_t}_{n_t}]^\sharp)
-(q^{n_{t-1}+1}_{j_{t-1}}|q^{j_t}_{n_t}) (\lambda+\partial)
+z(s|[q^{n_{t-1}+1}_{j_{t-1}},q^{j_t}_{n_t}])
\big)
} \\
\displaystyle{
\phantom{\Big(}
\times\big(
w([q^{n_t+1}_{j_t},a]^\sharp)
-(q^{n_t+1}_{j_t}|a) \lambda
+z(s|[q^{n_t+1}_{j_t},a])
\big)
\,.}
\end{array}
\end{equation}
\end{theorem}
Note that in each summand of \eqref{20140304:eq4} 
the $z$ term can be non-zero at most in one factor.
In fact, $z$ may occur in the first factor only for $k_1\leq0$,
in the second factor only for $k_1\geq1$ and $k_2\leq-1$,
in the third factor only for $k_2\geq1$ and $k_3\leq-1$,
and so on, and it may occur in the last factor only for $k_t\geq1$.
Since these conditions are mutually exclusive,
the expression in the RHS of \eqref{20140304:eq4} is linear in $z$.

Some special cases and applications of equation \eqref{20140304:eq4} are summarized in the next result.
\begin{proposition}[\cite{DSKV16}]\phantomsection\label{20160203:prop1}
\begin{enumerate}[(a)]
\item
If either $a$ or $b$ lies in $\mf g^f_0$, we have
\begin{equation}\label{20140225:eq1}
\{{w(a)}_\lambda{w(b)}\}_{z,\rho}
=
w([a,b])+(a|b)\lambda+z(s|[a,b])
\,.
\end{equation}
\item
If $a,b\in\mf g^f_{-\frac12}$ we have
\begin{equation}\label{20140226:eq8}
\begin{array}{l}
\displaystyle{
\phantom{\Big(}
\{{w(a)}_\lambda{w(b)}\}_{z,\rho}
=
w([a,b])
+(\partial+2\lambda)
w([a,[e,b]]^\sharp)
-(e|[a,b])\lambda^2
} \\
\displaystyle{
\,\,\,\,\,\,\,\,\, \,\,\,\,\,\,\,\,\, \,\,\,\,\,\,\,\,\,
+\sum_{(j,n)\in J_{-\frac12}}
w([a,q^{j}_{n}]^\sharp)
w([q^{n+1}_{j},b]^\sharp)
+z(s|[a,b])
\,.}
\end{array}
\end{equation}
\item
Consider the element
$L_0=\frac12\sum_{j\in J^f_0}w(q_j)w(q^j)\,\in\mc W\{2\}$.
Then, the element $L=w(f)+L_0\in\mc W\{2\}$ is a Virasoro element of $\mc W$,
and we have
\begin{equation}\label{20140228:eq4}
\{L_\lambda L\}_{z,\rho}
=
(\partial+2\lambda)L-(x|x)\lambda^3+2z(s|f)\lambda
\,.
\end{equation}
For $a\in\mf g^f_{1-\Delta}$ we have
\begin{equation}\label{20140228:eq5}
\{L_\lambda w(a)\}_{z,\rho}
=
(\partial+\Delta\lambda)w(a)
-\frac{(e|a)}2\lambda^3+z\Delta(s|a)\lambda
\,.
\end{equation}
In particular,
for $z=0$, all the generators $w(a),\,a\in\mf g^f$, of $\mc W$ are primary elements for $L$,
provided that $(e|a)=0$.
In other words, for $z=0$, $\mc W$ is an algebra of differential polynomials
generated by $L$ and $\ell-1$ primary elements with respect to $L$.
So, $\mc W$ is a PVA of \emph{CFT type} (cf. \cite{DSKW10}).

\end{enumerate}
\end{proposition}

\begin{remark}
Equations \eqref{20140225:eq1} and \eqref{20140226:eq8} and the definition of the Virasoro element $L$ in in
Proposition \ref{20160203:prop1}(c) are compatible with the analogous in \cite{DSKV14} where the classical affine $\mc W$-algebra
for minimal nilpotent elements has been explicitly described.
\end{remark}

\section{The package MasterPVA}\label{sec:3}
In this section we show how to use the package \textsl{MasterPVA}, both in its one- and multi-dimensional versions.
As a few examples, we prove the compatibility between GFZ and Virasoro-Magri PVA (case $N=D=1$),
we derive the Dubrovin-Novikov conditions for a bidimensional Poisson structure of hydrodynamic type
(case $D=1, N=2$) and we obtain the Mokhov's classification for the $N=1$ multidimensional structures of hydrodynamic type  \cite{M88}.

The packages \texttt{MasterPVA.m} and \texttt{MasterPVAmulti.m} must be in a directory where Mathematica can find them. This can be achieved, for example, by using the command \cmd{SetDirectory}. After this, we can load the packages. The two packages cannot be loaded in the same session, because of the conflicting functions and properties definition. However, MasterPVAmulti can effectively deal with $D=1$ PVA, despite using a heavier notation. This is the reason why we provide a package specifically devoted to standard monodimensional PVAs, althought the same input works with \textsl{MasterPVAmulti}.
\begin{flushleft}
\hspace{1mm}\includegraphics[scale=0.55]{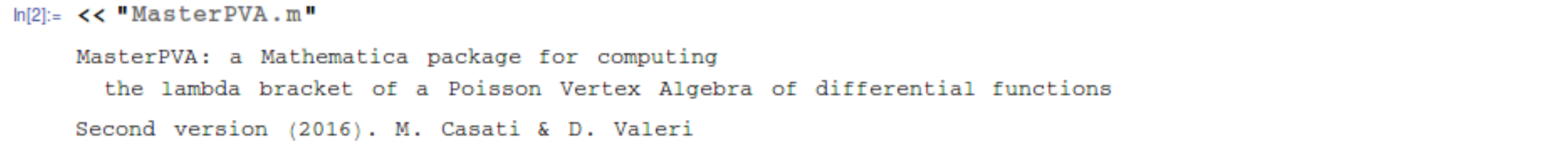}
\end{flushleft}

\subsection{GFZ and Virasoro-Magri Poisson vertex algebras}\label{sec:kdv}
Let $\mc V$ be an algebra of differential functions extending $R_1=\mb C[u,u',u'',\ldots]$. We recall that the
Gardner-Faddeev-Zacharov (GFZ) PVA structure on $\mc V$ is defined by
\begin{equation}\label{eq:GFZdef}
 \{u_\lambda u\}_1=\lambda
\end{equation}
while the Virasoro-Magri PVA with central charge $c\in\mb C$ is defined by
\begin{equation}\label{eq:VMdef}
 \{u_\lambda u\}_0=\left(\dev+2\lambda\right)u+c\lambda^3\,.
\end{equation}
We will show the well-known fact that these two structures are compatible, namely that the $\lambda$-bracket 
$\{\cdot\,_\lambda\,\cdot\}_z=\{\cdot\,_\lambda\,\cdot\}_0+z\{\cdot\,_\lambda\,\cdot\}_1$ defines
a PVA structure on $\mc V$ for all $z\in\mb C$.

After loading the package, it is necessary to set the number of generators, the name for the generators, for the independent variable with respect to which the derivation $\dev$ acts, and for the formal indeterminate used in the definition of the $\lambda$-brackets, say $\lambda$. The syntax for these commands is
\begin{flushleft}
\hspace{1mm}\includegraphics[scale=0.55]{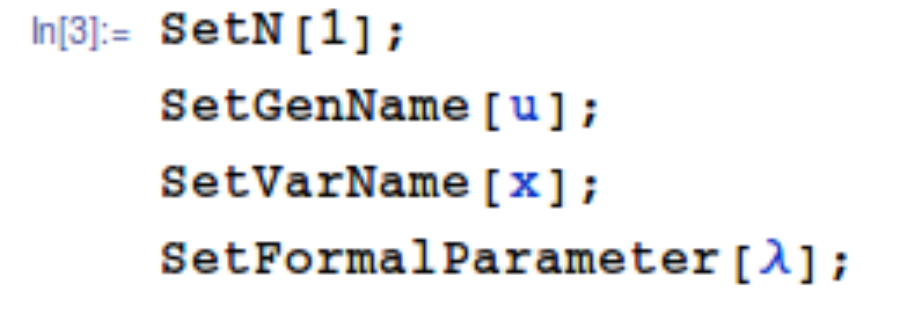}
\end{flushleft}
The list of generators, written as functions of the independent variables, is called \cmd{gen} throughout the program. The $\lambda$-brackets between the generators must be provided in form of a $N\times N$ table, whose entries are polynomials in the previously declared formal indeterminate. In this example $N=1$ and we have \cmd{H0} given by equation \eqref{eq:VMdef}
 and \cmd{H1} given by equation \eqref{eq:GFZdef}. We denote by \cmd{H} their linear combination.
\begin{flushleft}
\hspace{1mm}\includegraphics[scale=0.55]{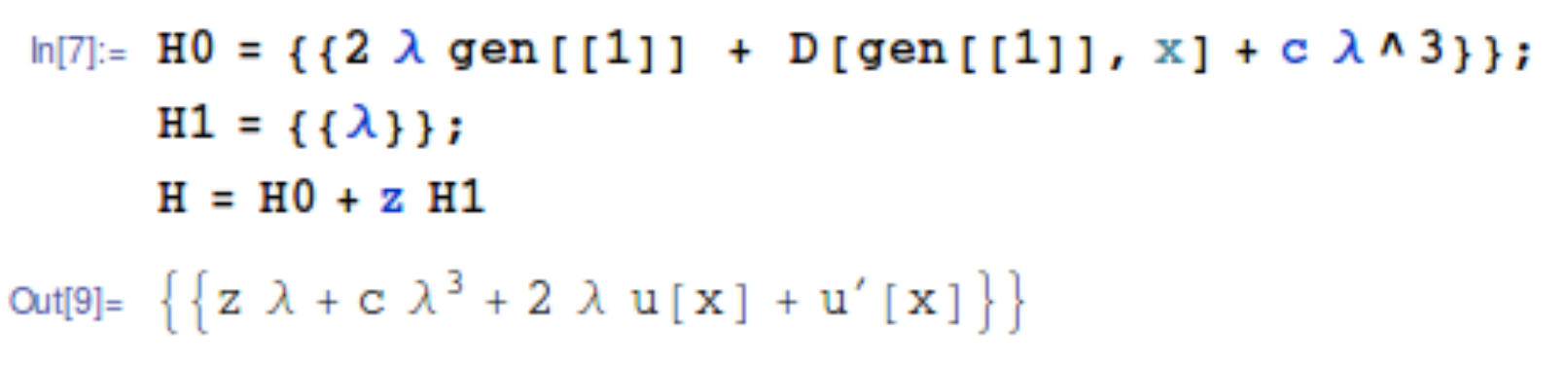}
\end{flushleft}
The skewsymmetry and Jacobi identity on generators (see Theorem \ref{master}) can be checked by using the functions \cmd{PVASkew[]} and \cmd{JacobiCheck[]}.
Indeed the output of \cmd{PVASkew[]} (respectively \cmd{JacobiCheck[]}) is the LHS of 
equation \eqref{skewsimgen} (respectively \eqref{jacobigen}).
We get
\begin{flushleft}
\hspace{1mm}\includegraphics[scale=0.55]{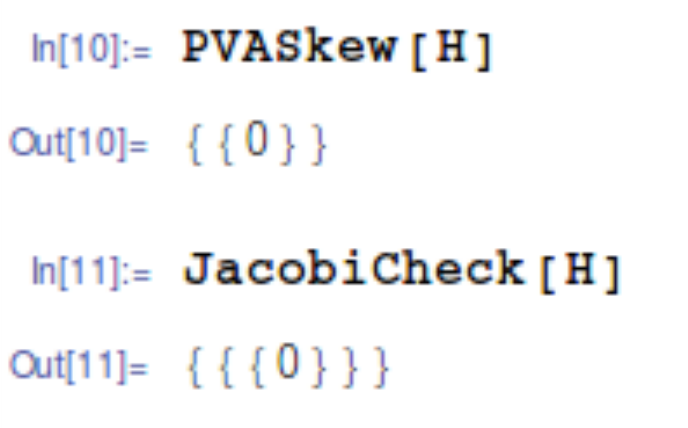}
\end{flushleft}
thus showing that $H_0$ and $H_1$ define two compatible PVA structures on $\mc V$.

Let us define $h_1=\frac12 u^2\in\mc V$.
It is well known that the corresponding Hamiltonian equation \eqref{hameq}
corresponding to the Hamiltonian functional $\tint h_1$ and the Poisson structure
\cmd{H0} is the Korteweg-de Vries (KdV) equation.
Moreover, let us also define $h_2=\frac12u^3+\frac c2uu''\in\mc V$.
The KdV equation is also the Hamiltonian equation corresponding to the Hamiltonian functional
$\tint h_2$ and the Poisson structure \cmd{H1}. 
\begin{flushleft}
\hspace{1mm}\includegraphics[scale=0.55]{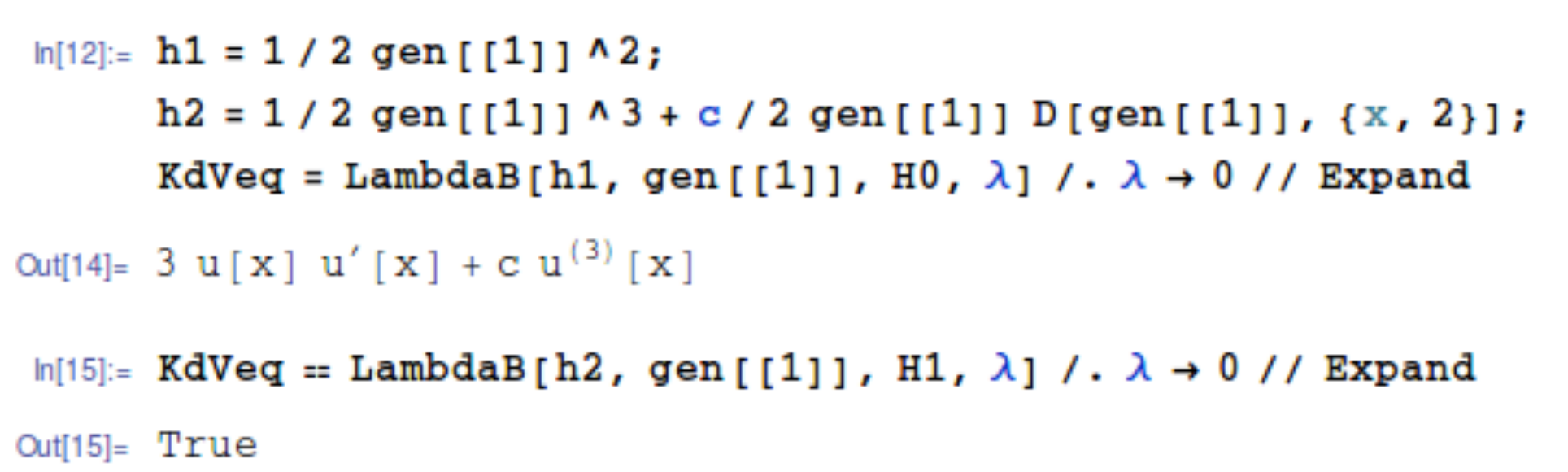}
\end{flushleft}
In fact, the KdV equation is a bi-Hamiltonian equation and its integrability can be proved using 
the Lenard-Magri scheme of integrability \cite{Mag78}.

\subsection{Poisson structures of hydrodynamic type}
Let $\mc V$ be an algebra of differential functions extending $R_N$.
A Poisson structure of hydrodynamic type \cite{DN83} on $\mc V$  is defined by the following $\lambda$-bracket on
generators ($i,j,k=1,\dots,N$):
\begin{equation}\label{eq:HYPB}
 \{u_i{}_\lambda u_j\}=g_{ji} \lambda+b_{ji}^k u'_k\,,
\end{equation}
where repeated indices are summed according to Einstein's rule and
$\frac{\partial g_{ji}}{\partial u_h^{(n)}}=\frac{\partial b_{ji}^k}{\partial u_h^{(n)}}=0$, for every $h=1,\dots,N$ and $n\geq1$.

The geometric interpretation of the functions $g_{ij}$ and $b_{ij}^k$ is well known:
the $\lambda$-bracket defined in \eqref{eq:HYPB} defines a PVA structure on $\mc V$ if and only $g_{ij}$ are the components of a flat contravariant metric on a manifold with local coordinates $(u_1,\ldots,u_N)$ and $b_{ij}^k$ are the contravariant Christoffel symbols of the associated Levi-Civita connection. Using \textsl{MasterPVA} we will derive the explicit form of these properties in the case $N=2$.

After loading the package, we initialize the package settings.
\begin{flushleft}
\hspace{1mm}\includegraphics[scale=0.55]{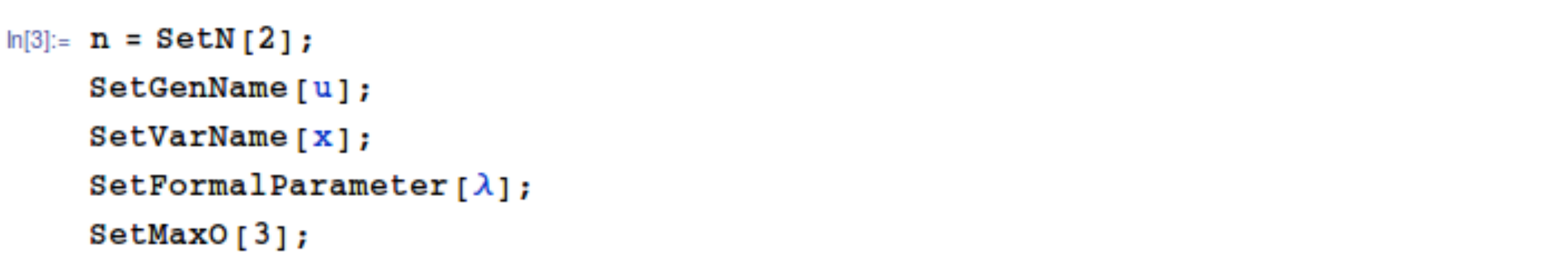}
\end{flushleft}
We define the matrices $g_{ij}$ and $b_{ij}^k$ and use them to write the $\lambda$-bracket \eqref{eq:HYPB}.
\begin{flushleft}
\hspace{1mm}\includegraphics[scale=0.50]{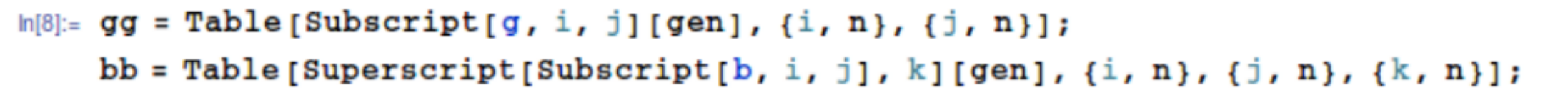}
\hspace{1mm}\includegraphics[scale=0.50]{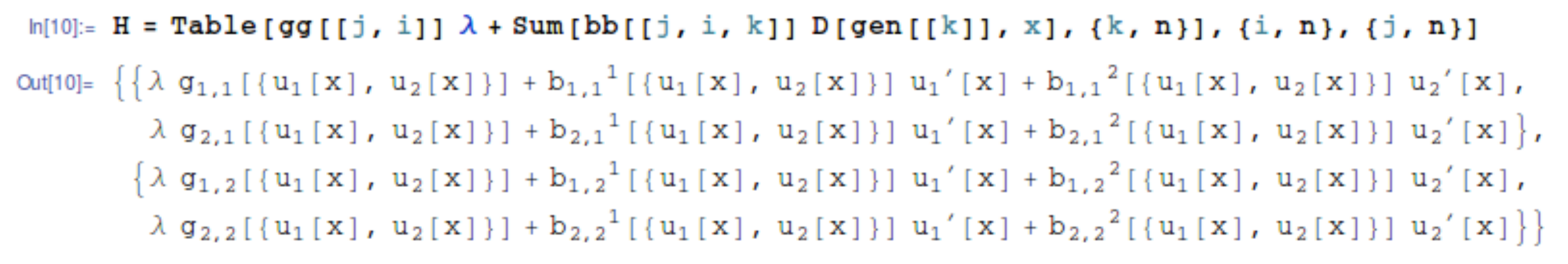}
\end{flushleft}
By equating to zero the coefficient of $\lambda$ and the constant term (in $\lambda$) in the equations given by \texttt{PVASkew[P]}
we get the conditions that $g_{ij}$ and $b_{ij}^k$ should satisfy in order to get a skewsymmetric $\lambda$-bracket.
\begin{flushleft}
\hspace{1mm}\includegraphics[scale=0.50]{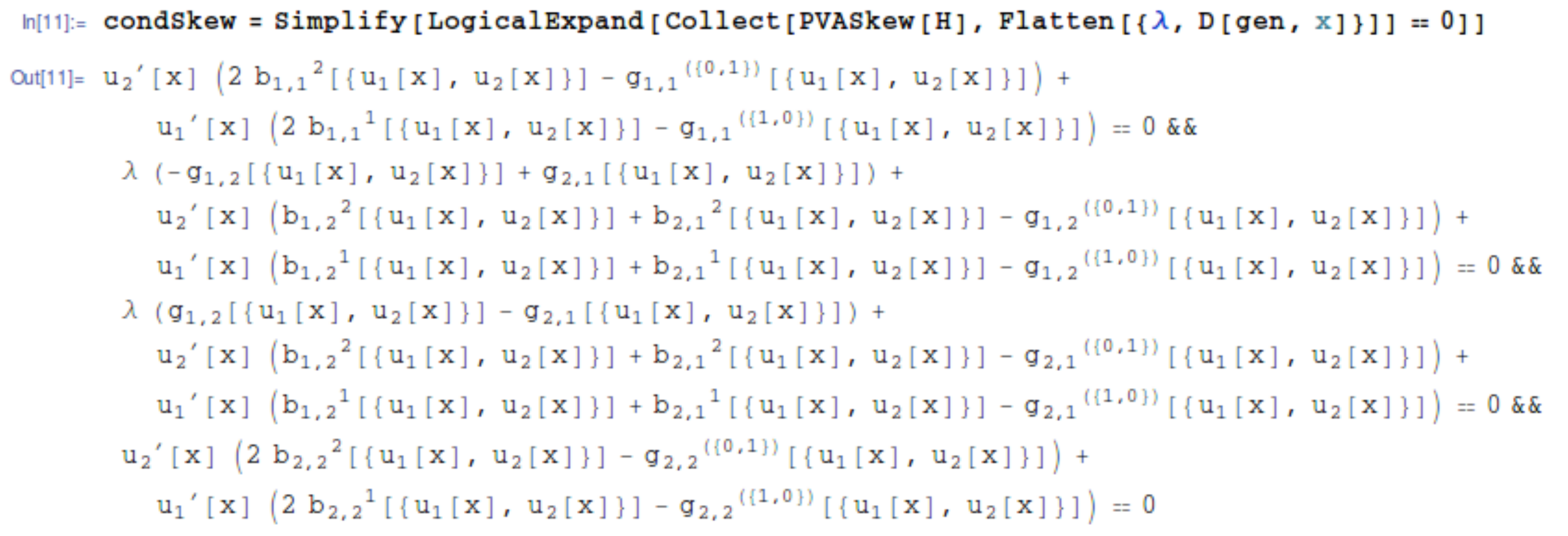}
\end{flushleft}
These conditions can be summarized by the equations
\begin{equation}\label{eq:HYPBskew}
 g_{ij}=g_{ji}\,,
 \qquad
 b_{ij}^k+b_{ji}^k=\frac{\dev g_{ij}}{\dev u_k}\,.
\end{equation}
We redefine the functions $g_{ij}$ and $b_{ij}^k$ and the Poisson structure \texttt{H} in order to ensure the validity of
equations \eqref{eq:HYPBskew}.
\begin{flushleft}
\hspace{1mm}\includegraphics[scale=0.50]{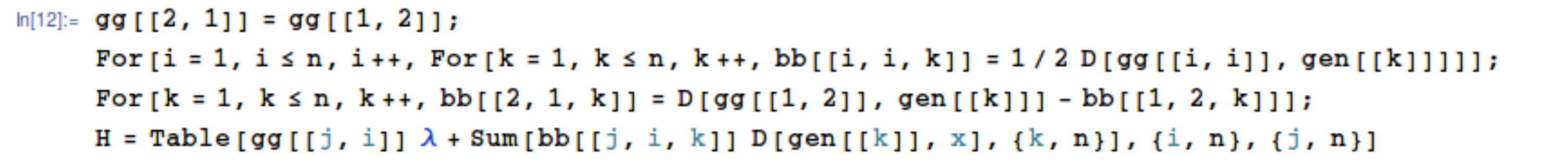}
\end{flushleft}
The further properties that must be satisfied to grant the Jacobi identity can be found using \texttt{JacobiCheck[P]}. Notice that, when the result of \texttt{JacobiCheck[]} is not identically vanishing, the output uses internal variables whose name starts with \texttt{MasterPVA`Private`}: to make the output clearer it is advisable to replace them with the ``external'' names, as it is demonstrated in the following picture. However, reading the conditions for the Jacobi identity is usually much more cumbersome than inspecting the ones for the skewsymmetry.
\begin{flushleft}
\hspace{1mm}\includegraphics[scale=0.50]{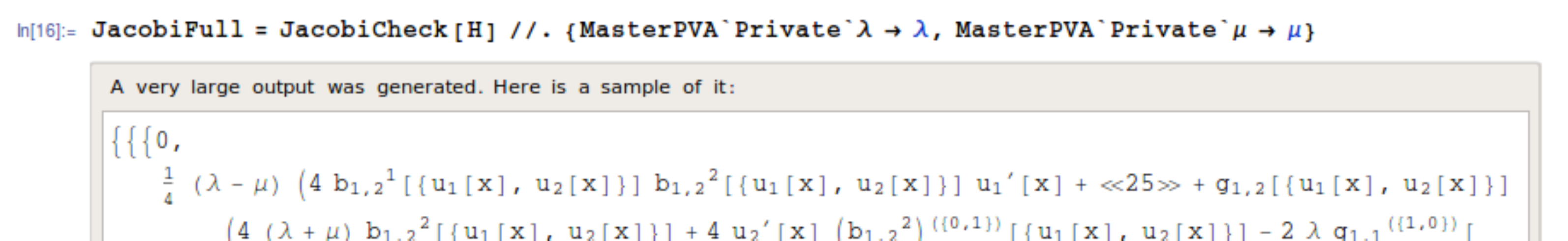}
\end{flushleft}
Nevertheless, we can check that the vanishing of the coefficient of $\lambda^2$ in the Jacobi identity is equivalent to the torsion--free condition for the Levi--Civita connection:
\begin{equation}
 g_{ia}b_{kj}^a-g_{ja}b_{ki}^a=0
 \,.
\end{equation}
\begin{flushleft}
\hspace{1mm}\includegraphics[scale=0.45]{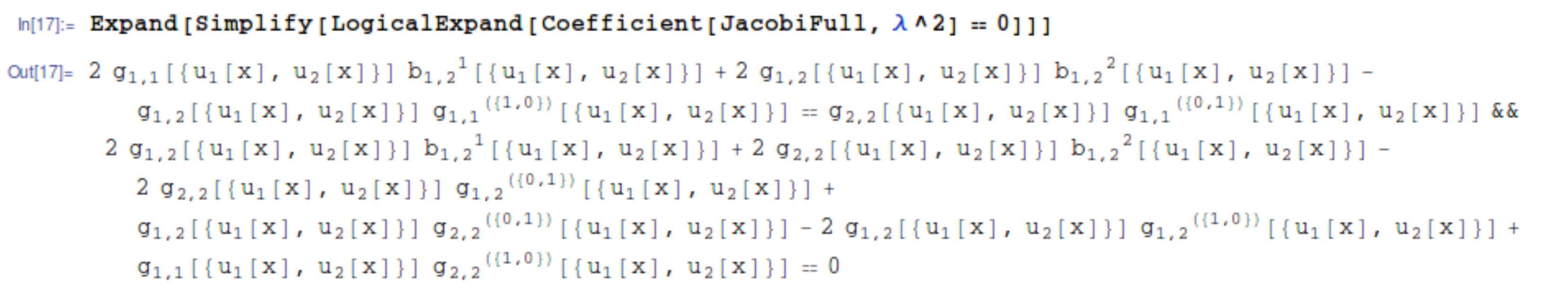}\\
\hspace{1mm}\includegraphics[scale=0.45]{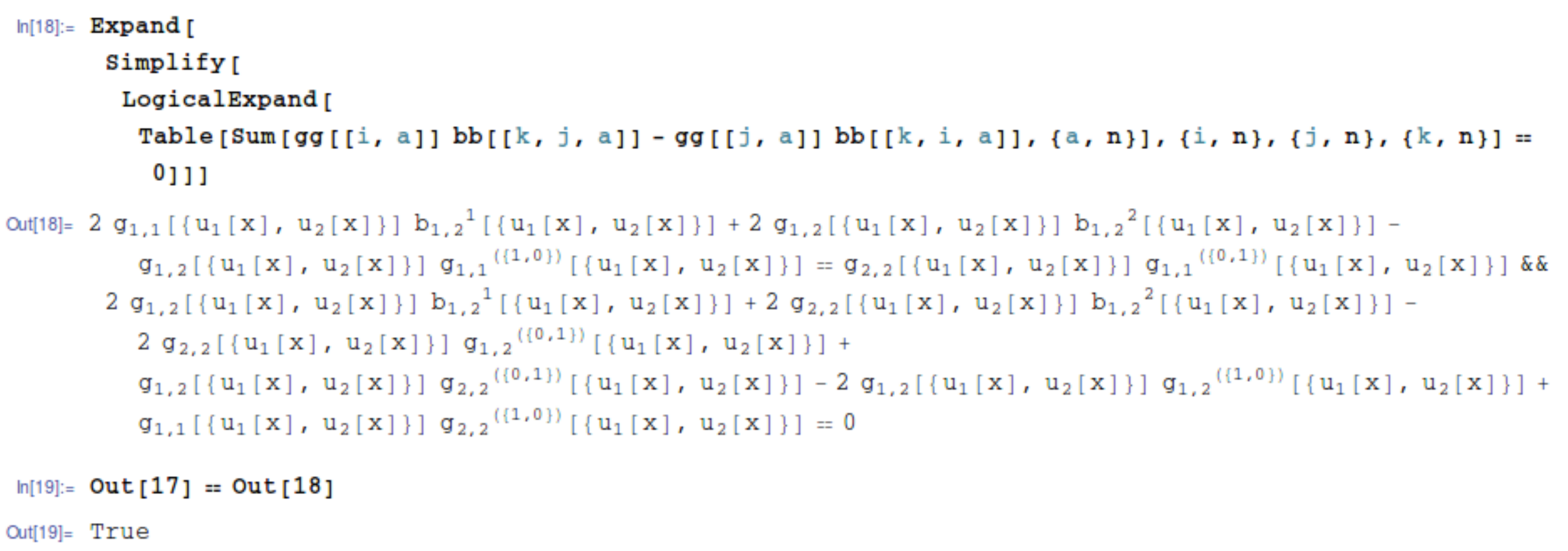}\\
\end{flushleft}

\subsection{Multidimensional scalar PVAs of hydrodynamic type}
The package \textsl{MasterPVAmulti} must be used when dealing with multidimensional PVA
defined in Section \ref{sec:multi}.
Here, we use it to classify multidimensional Poisson structures of hydrodynamic type
for the case $N=1$, $D=3$. This is a special case of a classification theorem proved by
Mokhov, \cite{M88}.

We recall that a multidimensional scalar
$\lambda$-bracket of hydrodynamic type has the form
\begin{equation}\label{20160212:eq1}
\{u_\lambda u\}=\sum_{\alpha=1}^D \left(a_\alpha\lambda_\alpha+ b_\alpha u_\alpha\right)
\,,
\end{equation}
where we set $u_\alpha=\dev_\alpha u$ and $a_\alpha$ and $b_\alpha$ are such that
$\frac{\partial a_\alpha}{\partial \partial^n_\beta u}
=\frac{\partial b_\alpha}{\partial \partial^n_\beta u}=0$, for every $\beta=1,\dots,D$ and $n\geq1$.

Mokhov's theorem states that the $\lambda$-bracket \eqref{20160212:eq1} defines a PVA structure if and only if it is of the form
\begin{equation}\label{eq:mHYPB}
 \{u_\lambda u\}=\sum_{\alpha=1}^D c_\alpha\left(2g\lambda_\alpha+ u_\alpha\frac{\partial g}{\partial u} \right)
 \,,
\end{equation}
for some $c_\alpha\in\mb C$ and a function $g$
such that $\frac{\partial g}{\partial \partial^n_\beta u}=0$,
for every $\beta=1,\dots,D$ and $n\geq1$.

After loading the package we initialize the variables similarly to what we did at the beginning of
Section \ref{sec:kdv}, but in this case we should specify the spatial dimension $D$.
\begin{flushleft}
\hspace{1mm}\includegraphics[scale=0.55]{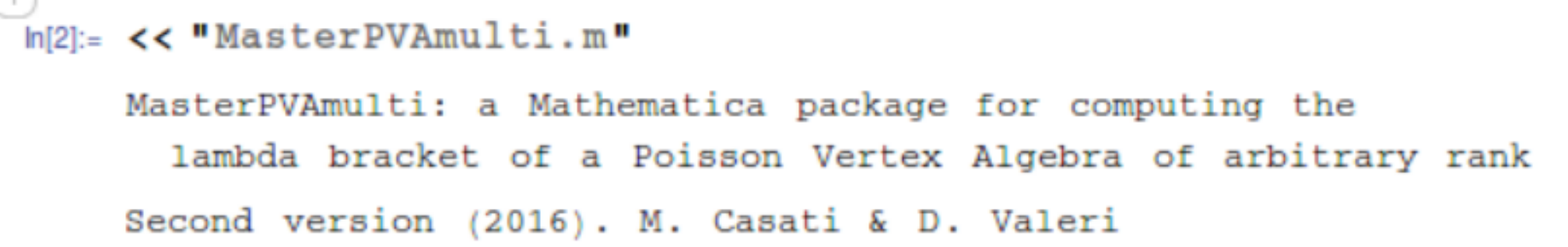}\\
\hspace{1mm}\includegraphics[scale=0.55]{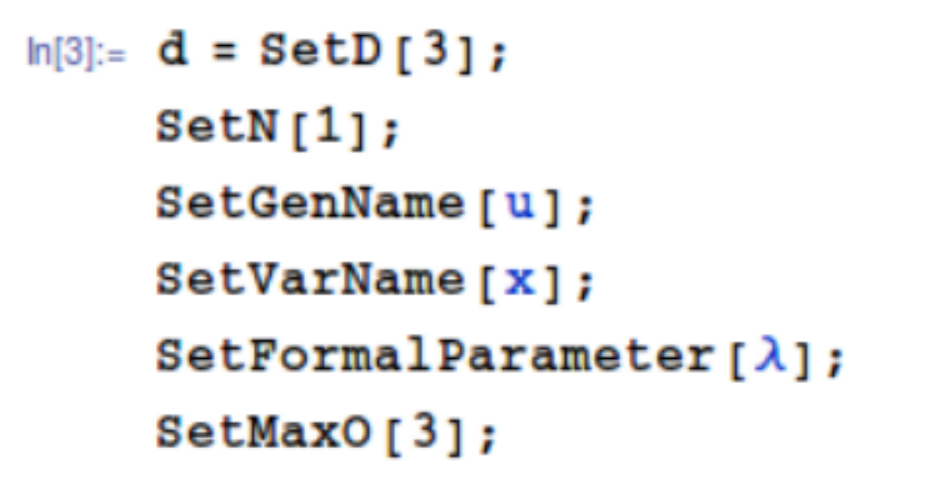}
\end{flushleft}
We define the $\lambda$-bracket as in equation \eqref{20160212:eq1} assuming $N=1$ and $D=3$.
The formal parameter, for which we chose the symbol $\lambda$ in the initialization, is $\{\lambda_1,\lambda_2,\lambda_3\}$.
\begin{flushleft}
\hspace{1mm}\includegraphics[scale=0.55]{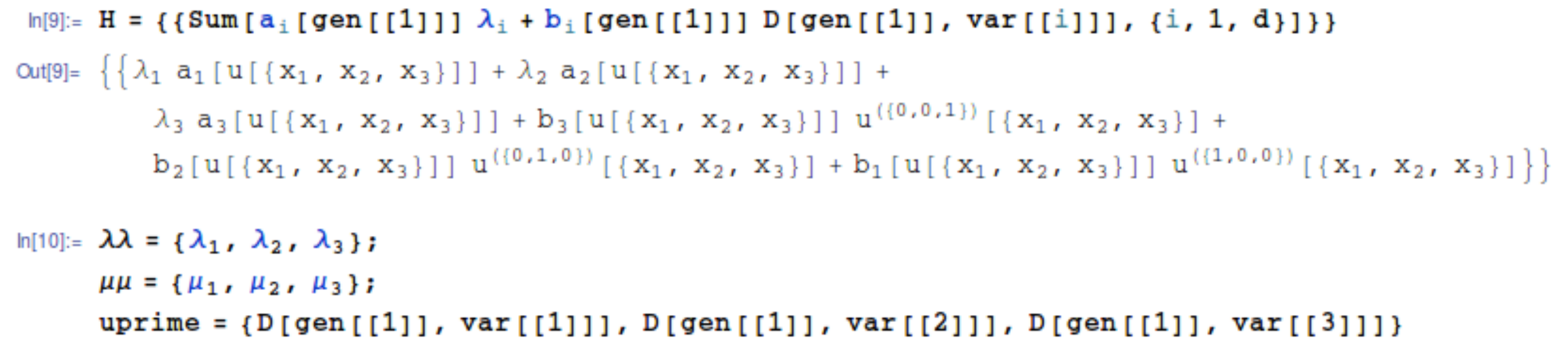}
\end{flushleft}
We use \cmd{PVASkew[P]} to find the conditions that the functions $a_\alpha$ and  $b_\alpha$ should satisfy in order to have
a skewsymmetric $\lambda$-bracket. We get
\begin{equation}\label{eq:mHYPB_skew}
2 b_\alpha =\frac{\partial a_\alpha}{\partial u}
\,. 
\end{equation}
Hence, we define a new $\lambda$-bracket, called \cmd{Hskew}, where the functions $a_\alpha$ and $b_\alpha$ satisfy equation
\eqref{eq:mHYPB_skew}. Note that \cmd{Hskew} only depends on the functions $a_\alpha$ now.
\begin{flushleft}
\hspace{1mm}\includegraphics[scale=0.55]{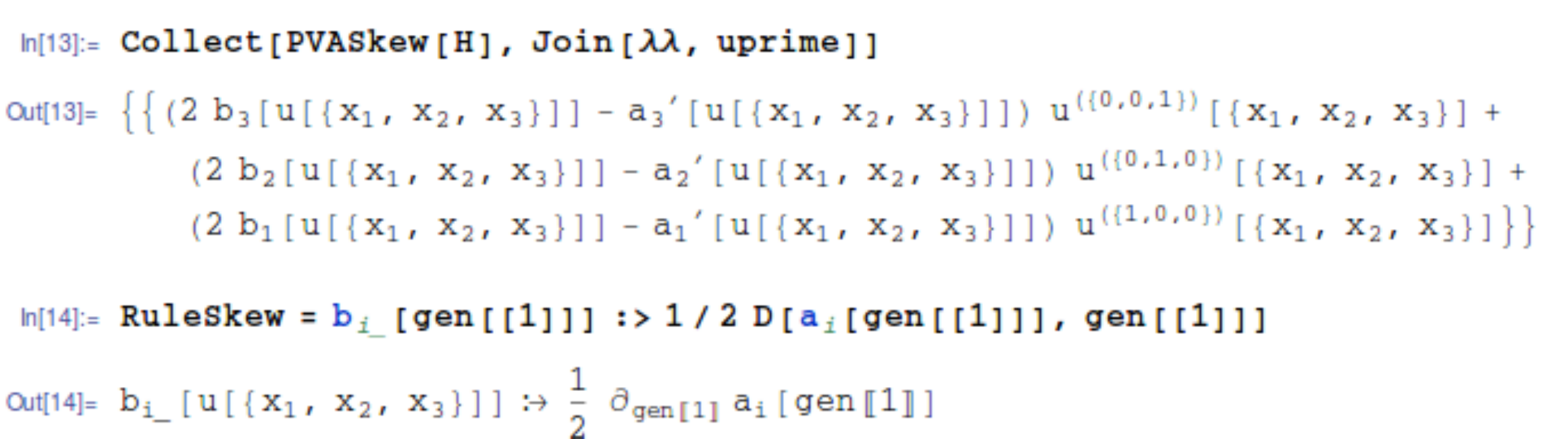}\\
\hspace{1mm}\includegraphics[scale=0.55]{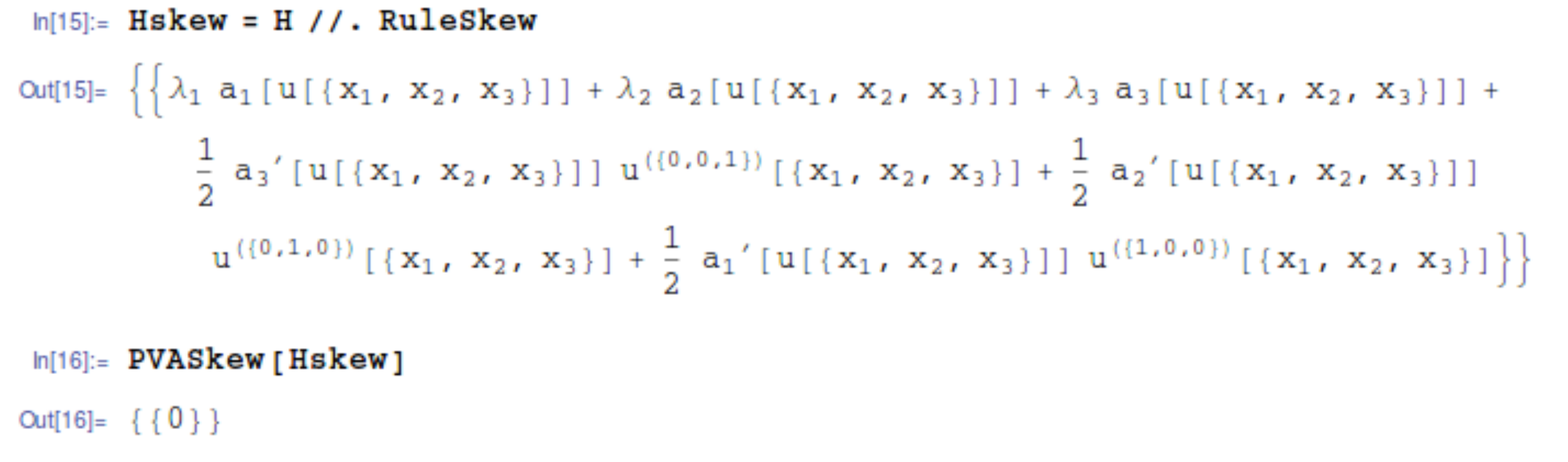}\\
\end{flushleft}
We use \cmd{JacobiCheck[Pskew]} to write the conditions that must be satisfied by the functions $a_\alpha$
in order to get the validity of \eqref{jacobigen}. We denote the LHS of \eqref{jacobigen} by \cmd{JacobiCond}. In particular, by equating to zero the coefficient of $\lambda_\alpha\mu_\beta$ we get a system of ODEs for the functions $a_\alpha$.
\begin{flushleft}
\hspace{1mm}\includegraphics[scale=0.55]{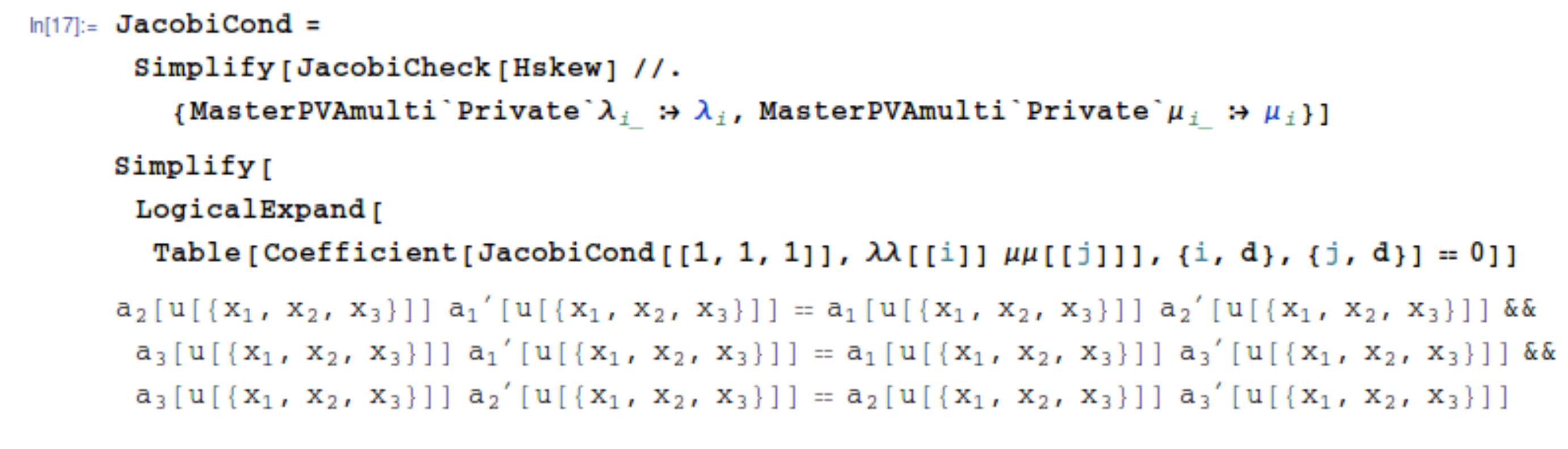}
\end{flushleft}
A solution to this system is given by
\begin{equation}\label{eq:mHYPB_sol}
 a_\alpha=c_\alpha g \qquad\qquad \text{for some function }g\text{ and }c_\alpha\in\mb C
 \,.
\end{equation}
Then, we can substitute equation \eqref{eq:mHYPB_sol} in \cmd{JacobiCond}.
\begin{flushleft}
\hspace{1mm}\includegraphics[scale=0.55]{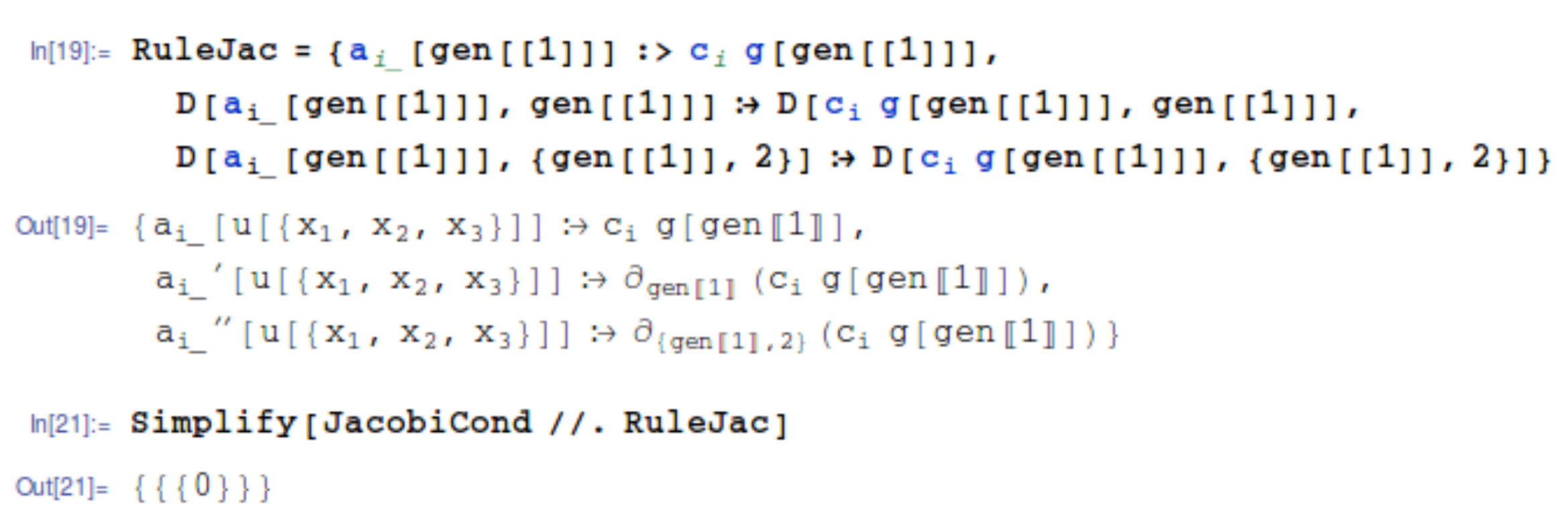}
\end{flushleft}
Hence, Jacobi identity \eqref{jacobigen} holds, thus showing that \eqref{eq:mHYPB_skew} and \eqref{eq:mHYPB_sol}
are the sufficient and necessary conditions for the $\lambda$-bracket \eqref{20160212:eq1} to define a PVA structure,
as Mokhov's theorem states.

\section{The package WAlg}\label{sec:WAlg}
In this section we show how to use the package \textsl{WAlg}.
Its main function is to compute the $\lambda$-brackets between the generators of the classical affine $\mc W$-algebra
$\mc W(\mf g,f)$ defined in Section \ref{sec:2.1}, where $\mf g$ is a  simple Lie algebra of type $A,B,C,D$ and $G$, and $f\in\mf g$ is a nilpotent element.
Thus, we can use this result to compute the $\lambda$-brackets between arbitrary elements of the classical $\mc W$-algebra
and the corresponding generalized Drinfeld-Sokolov hierarchies using the package \textsl{MasterPVA}.
%

In order to perform computations with \textsl{WAlg} we need to realize the simple Lie algebras
of type $A,B,C,D$ and $G$ as subalgebras of $\mf{gl}_N$.
(We emphasize that we can do the same for simple Lie algebras of 
type $E$ and $F$. Unfortunately the dimension of such a representation can be big, as for the case of $E_8$, where the minimal $N=248$.)

Given an element $A=(A_{ij})_{i,j=1}^N\in\mf{gl}_N$, we denote by $A^\text{at}=\left((A^{\text{at}})_{ij}\right)_{i,j=1}^N$ its transpose with respect to the antidiagonal, namely
$(A^{\text{at}})_{ij}=A_{N+1-j,N+1-i}$. Then, we realize the classical Lie algebras as in \cite{DS85}:
\begin{enumerate}[(A)]
\item Type $A_n$: $\mf g=\mf{sl}_n=\{A\in\mf{gl}_{n+1}\mid \tr(A)=0\}$.
\item Type $B_n$: let $S=\sum_{k=1}^{2n+1}(-1)^{k+1}E_{kk}$, then
$\mf g=\mf o_{2n+1}=\{A\in\mf{gl}_{2n+1}\mid A=-SA^{\text{at}}S\}$.
\item Type $C_n$: let $S=\sum_{k=1}^{2n}(-1)^{k+1}E_{kk}$, then
$\mf g=\mf{sp}_{2n}=\{A\in\mf{gl}_{2n}\mid A=-SA^{\text{at}}S\}$.
\item Type $D_n$: let $S=\sum_{k=1}^{n}(-1)^{k+1}(E_{kk}+E_{2n+1-k,2n+1-k})$, then
$\mf g=\mf{o}_{2n}=\{A\in\mf{gl}_{2n}\mid A=-SA^{\text{at}}S\}$.
\end{enumerate}
In the sequel, given a matrix $A\in\mf{gl}_N$, we denote by $\sigma(A)=-SA^{\text{at}}S$, where $S$ can be any of the matrix
appearing in the definition of the classical Lie algebras of type $B,C$ and $D$. Clearly, $A+\sigma(A)$ belongs to the 
corresponding classical Lie algebra, since $\sigma^2=\id_{\mf{gl}_N}$.

We realize $G_2$ as a subalgebra of $D_4$ as follows. Note that the group of automorphisms
of the Dynkin diagram of $D_4$ is isomorphic to $S_3$, the group of permutations on three elements. Then, we can consider the induced action by Lie algebra automorphisms of this group on $\mf o_8$.  Then, it is easy to check that:
\begin{enumerate}[(G)]
\item Type $G_2$: $\mf g=\{A\in\mf o_8\mid \tau(A)=A\,,\text{for every }\tau\in S_3\}$.
\end{enumerate}
In particular, we used the following choice of Chevalley generators for $\mf g$:
\begin{align*}
&e_1=E_{23}+E_{67}\,,& &e_2=E_{12}+E_{34}+E_{56}+E_{78}+\frac{1}{2}\left(E_{35}+E_{46}\right)\,,
\\
&h_1=E_{22}-E_{33}+E_{55}-E_{66}\,,& &h_2=E_{11}-E_{22}+2E_{33}-2E_{66}
+E_{77}-E_{88}\,,
\\
&f_1=E_{32}+E_{76}\,, & &f_2= E_{21}+E_{43}+E_{65}+E_{87}+2\left(E_{53}+E_{64}\right)
\,.
\end{align*}

After the choice of the simple Lie algebra $\mf g$ we need to choose a nilpotent element $f\in\mf g$. Since the construction
of classical affine $\mc W$-algebras does not depend on the nilpotent element itself, but only on its nilpotent orbit
(see \cite{DSKV13}), we assume that the nilpotent element is given in input as a strictly lower triangular matrix.
In fact, when giving in input a nilpotent element we can use the classification of nilpotent orbits given in \cite{CMG93}.
Then the program computes an $\mf{sl}_2$-triple $\{f,h=2x,e\}\subset\mf g$ such that $x$ is a diagonal matrix
and $e$ is strictly upper triangular.

Finally, we always assume that the nondegenerate symmetric invariant bilinear form on
$\mf g$ is a multiple of the trace form on matrices ($a,b\in\mf g)$:
$$
(a|b)=c\tr(ab)\,,\qquad c\in\mb C^*\,.
$$

\subsection{The algebraic setup}
The package \textsl{WAlg} requires the use of the default library
\texttt{listK\_6.txt}. Hence the files \texttt{WAlg.m} and \texttt{listK\_6.txt} must be in a folder where Mathematica can find them.
It is also possible to use a different library as described in Section \ref{sec:5.3} to which we refer
for the technical details.
To select the working folder of Mathematica, where it will look for them and the potential output files will be saved, one may use the command \cmd{SetDirectory["path"]}. An alternative method to load the package, different from the one shown in Section \ref{sec:3}, is using the command \cmd{Needs[]}.

Let us use the package \textsl{WAlg} to get the explcit set of generators of the classical 
affine $\mc W$-algebra $\mc W(\mf{o}_7,f)$, where $f$ is the principal nilpotent element \cite{DS85}.
We load the package and use the command \cmd{InitializeWAlg[]}.
Recall that $\mf o_7$ is a classical Lie algebra of type $B_3$.
\begin{flushleft}
\hspace{1mm}\includegraphics[scale=0.55]{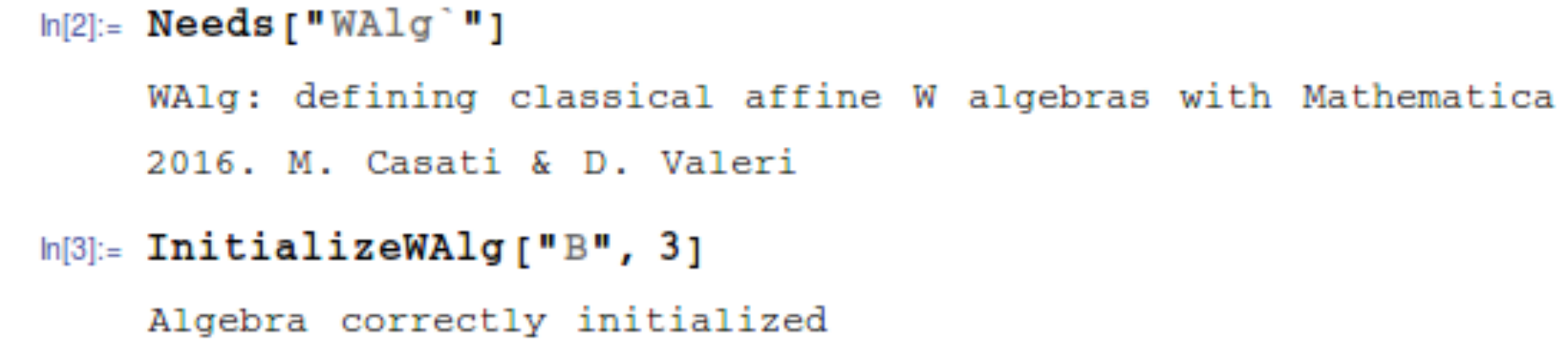}
\end{flushleft}
The dimension of the matrix representing $f$ is obtained with the command \cmd{GetDim[]}.
We define the principal nilpotent $f$, and we can also check that it belongs to $\mf o_7$.
The command \cmd{ComputeWAlg[]} takes the nilpotent element $f$ as argument and computes a basis of
$\mf g^f$ given by $\ad x$-eigenvectors. The warning notice is not a problem.
\begin{flushleft}
\hspace{1mm}\includegraphics[scale=0.55]{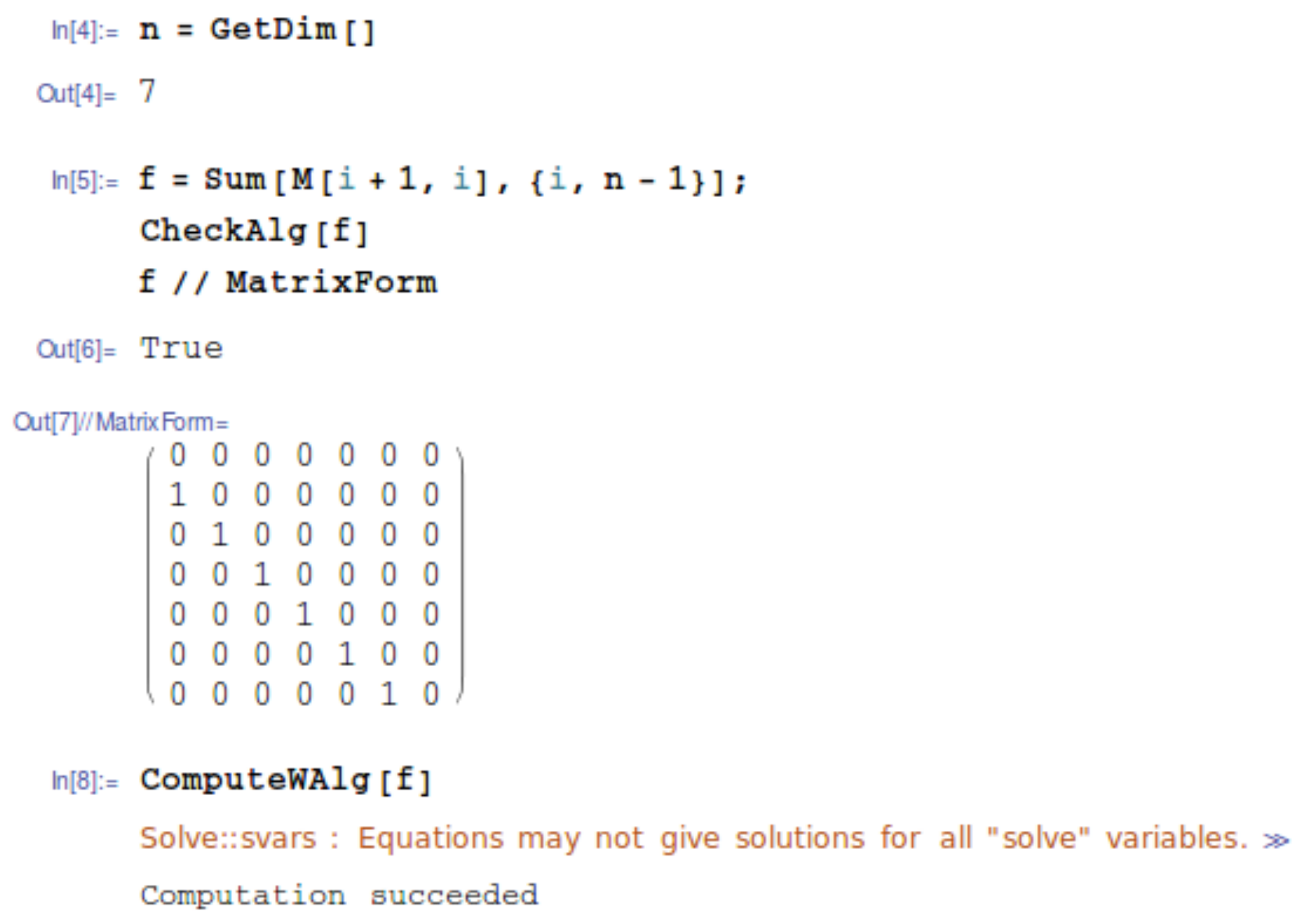}
\end{flushleft}
The basis computed for $\mf g^f$ can be recovered with the command \cmd{GetWBasis[]}, We denote it as \cmd{listq}.
The corresponding dual basis (with respect to the trace form) of $\mf g^e$ can be computed using the command \cmd{GetWbasisDual[]}. We denote it by
\cmd{listQ}. Finally, we can also use the command \cmd{GetWEigen[]} to recover all the $\ad x$-eigenvalues (with multiplicities) and put them in a list which we call \cmd{list$\delta$}.

The command \cmd{w[]} works as follows: it takes in input an element of $\mf o_7$, then it applies $\pi_{\mf g^f}$ and
the map $w$ defined in Theorem \ref{thm:structure-W} to this element. The result is a linear combination of the generators 
of the classical affine $\mc W$-algebra.
Hence, by Corollary \ref{20140221:cor}, \cmd{w[listq[[i]]]} gives the $i$-th generator of the classical affine $\mc W$-algebra.
(Note that, by an abuse of notation, these generators are denoted by $q_i$ in Mathematica, the same letter used to denote
the corresponding element of $\mf g^f$ to which they are attached through the map $w$. In fact, the notation $w(q_i)$ is used
in Corollary \ref{20140221:cor}.)
\begin{flushleft}
\hspace{1mm}\includegraphics[scale=0.55]{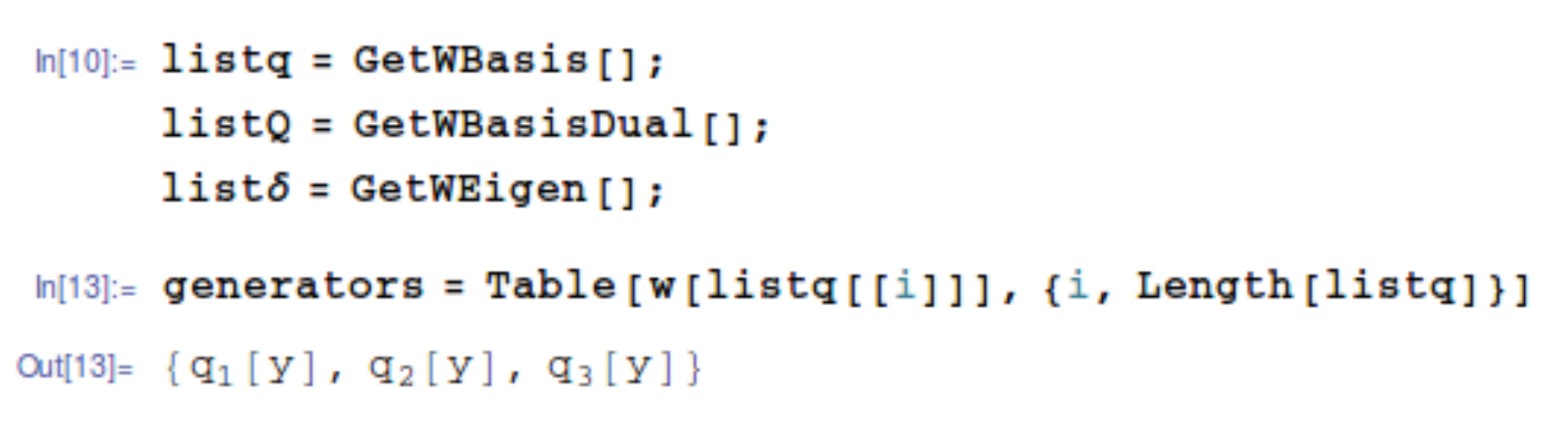}
\end{flushleft}
In the following example, we apply the command \cmd{w[]} to a random element of $\mf o_7$.
We construct it as follows: first we define a random element $A\in\mf{gl}_7$. Then,
using the function \cmd{Sigma[]} (which, given $A$ as input gives 
$\sigma(A)=-SA^{\text{at}}S$ as result) we get the element $A+\sigma(A)\in\mf o_7$
(we can also check it with the commad \cmd{CheckAlg[]}).

\begin{flushleft}
\hspace{1mm}\includegraphics[scale=0.55]{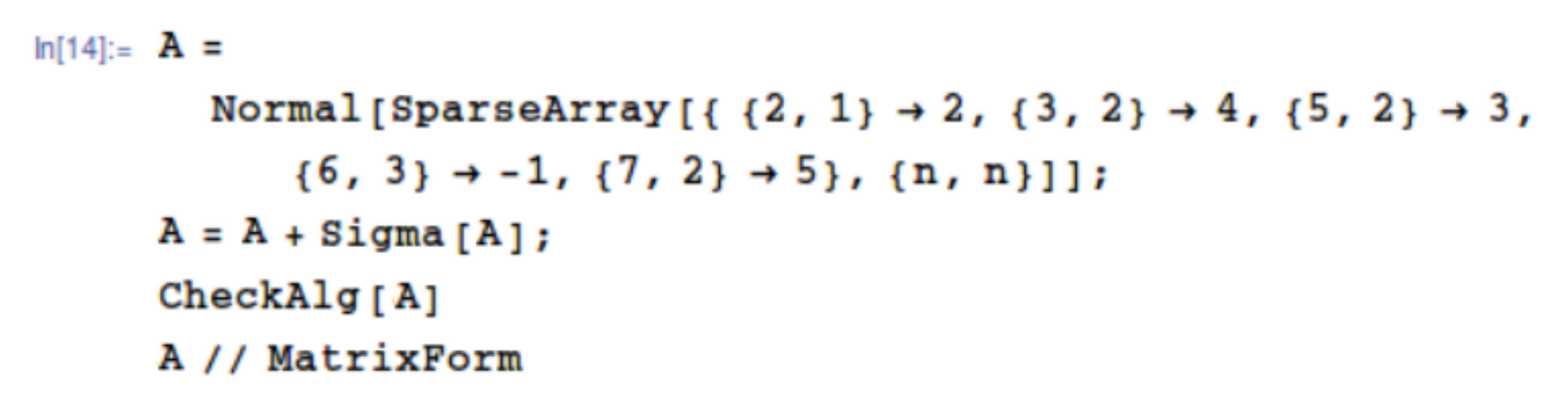}\\
\hspace{1mm}\includegraphics[scale=0.55]{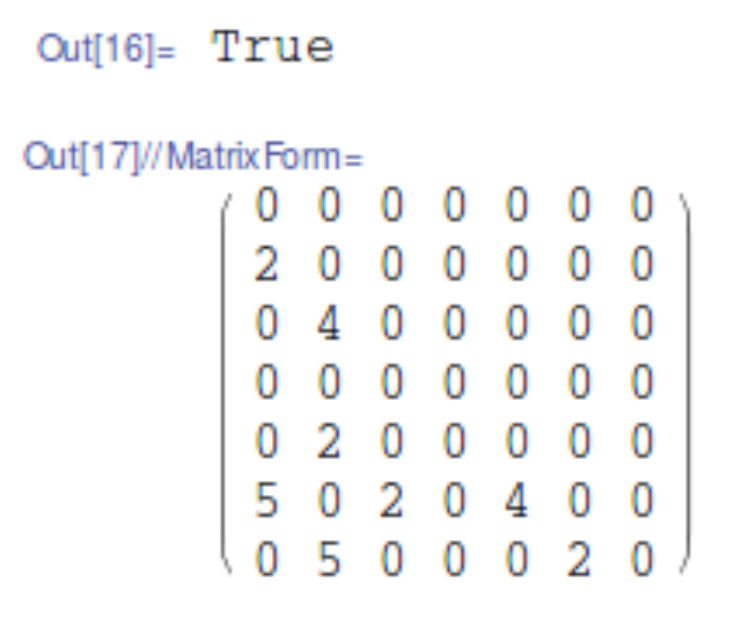}
\end{flushleft}
The command \cmd{w[]} gives the corresponding linear combination of the generators of the classical affine
$\mc W$-algebra, see Theorem \ref{thm:structure-W}.
\begin{flushleft}
\hspace{1mm}\includegraphics[scale=0.55]{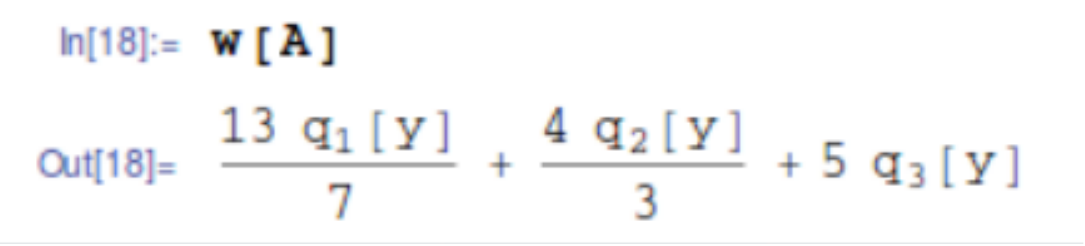}
\end{flushleft}
Note that, apart from computing explicitly the generators of the classical affine $\mc W$-algebra,
the command \cmd{w[]} is heavily used to implement equation \eqref{20140304:eq4}.
\subsection{Computation of $\lambda$-brackets among generators}
One of the most useful features of \textsl{WAlg} is the implementation of formula \eqref{20140304:eq4}
for the computation of the Poisson structure $H$, defined by equation \eqref{ham_W}, associated to
the classical affine $\mc W$-algebra.
After we compute $H$, we can use the package \textsl{MasterPVA} to compute the $\lambda$-brackets between
any elements of the classical affine $\mc W$-algebra.

Let us how how to proceed in the concrete example of the Lie algebra $\mf{sp}_4$ and its minimal nilpotent element $f$
\cite{DSKV14}.
Recall that $\mf{sp}_4$ is a classical Lie algebra of type $C_2$.
\begin{flushleft}
\hspace{1mm}\includegraphics[scale=0.55]{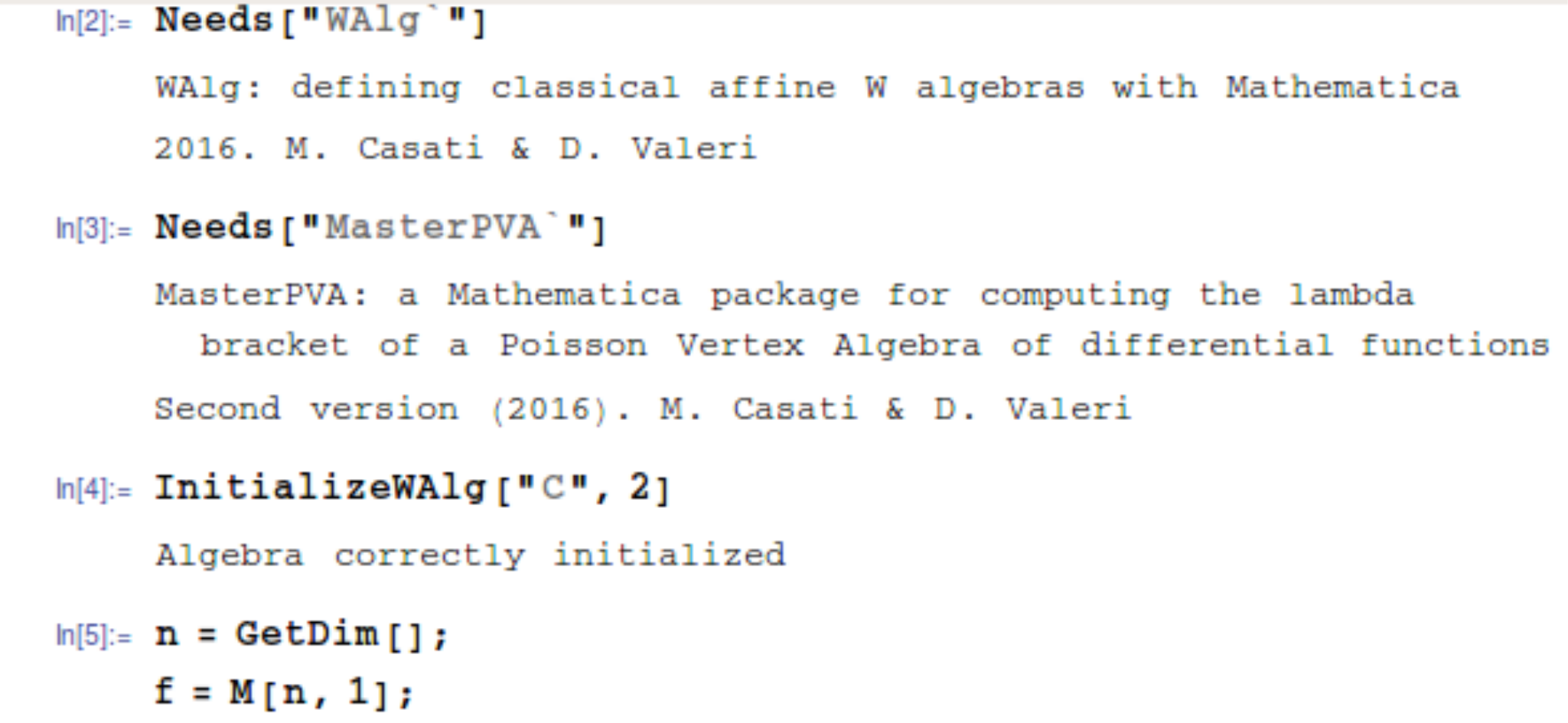}\\
\hspace{1mm}\includegraphics[scale=0.55]{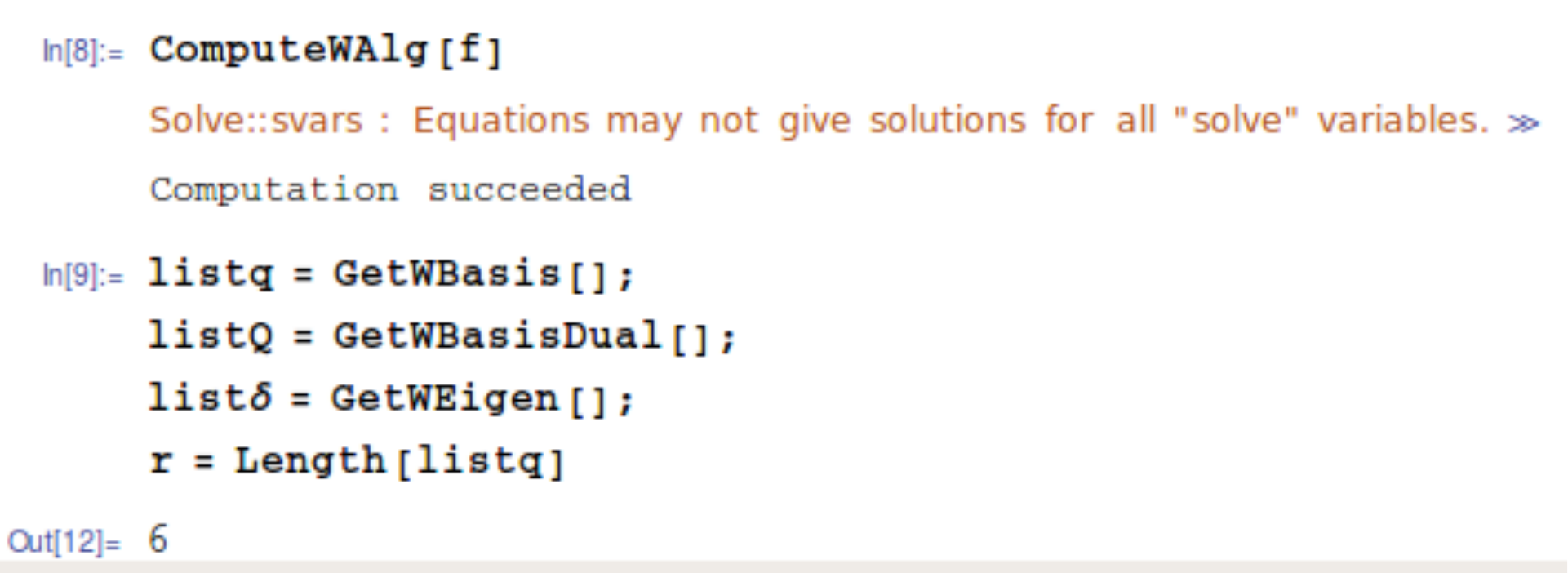}
\end{flushleft}
The number of generators of $\mc W(\mf{sp}_4,f)$ is the same as the dimension of $\mf g^f$,
which in this case is 6.
The command \cmd{SetS[]} allows us to set the element in $s\in\mf g_d$,
recall that $d$ is the maximal eigenvalue of $\ad x$, which is used in formula \eqref{20140304:eq4}.
If this command is left without argument it automatically choose a generic $s$. Note that in this example $\mf g_d=\mb Ce$,
so the choice is unique up to a constant.
Finally, the command \cmd{GenerateH[]} gives the Poisson structure $H$ associated to the classical affine $\mc W$-algebra
by equation \eqref{ham_W} implementing the formula \eqref{20140304:eq4}.
The optional parameter in the command is the formal parameter used in the definition of the $\lambda$-bracket, whose default value is $\beta$.
\begin{flushleft}
\hspace{1mm}\includegraphics[scale=0.55]{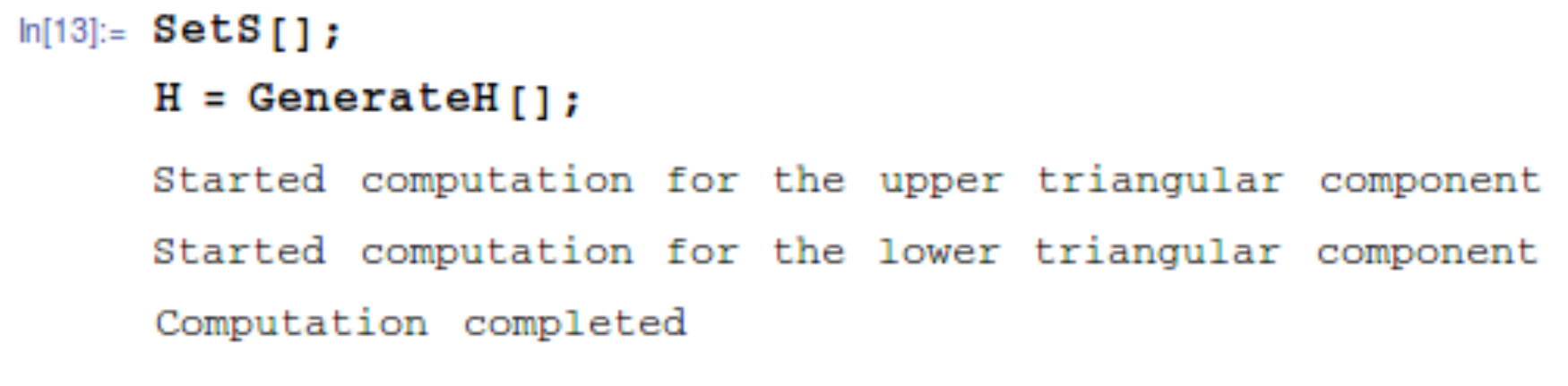}
\end{flushleft}
The next step is to allow the package \textsl{MasterPVA} to use the output of \cmd{GenerateH[]}.
In order to do that, we need to set the number of variables, use $q_i$ as the name of the generators,
use $y$ as independent variable, and use $\beta$ as the formal parameter.
\begin{flushleft}
\hspace{1mm}\includegraphics[scale=0.55]{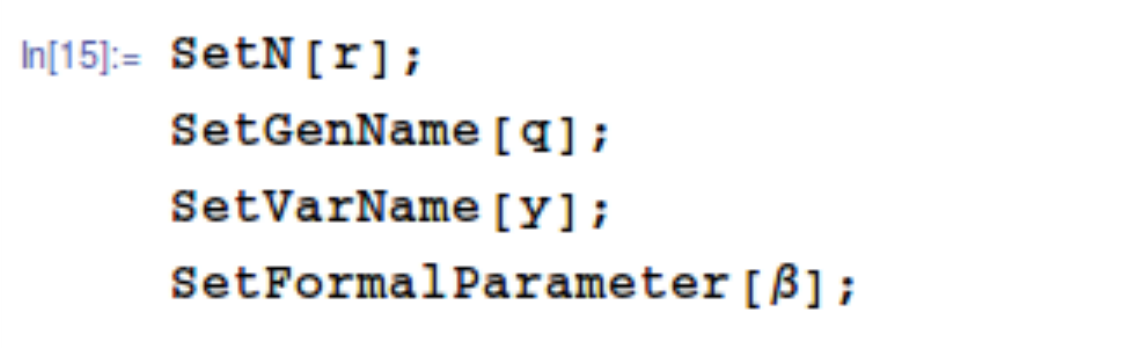}
\end{flushleft}
Now the commands of \textsl{MasterPVA} can be used. For example, we can check
that $H$ is indeed a Poisson structure.
\begin{flushleft}
\hspace{1mm}\includegraphics[scale=0.55]{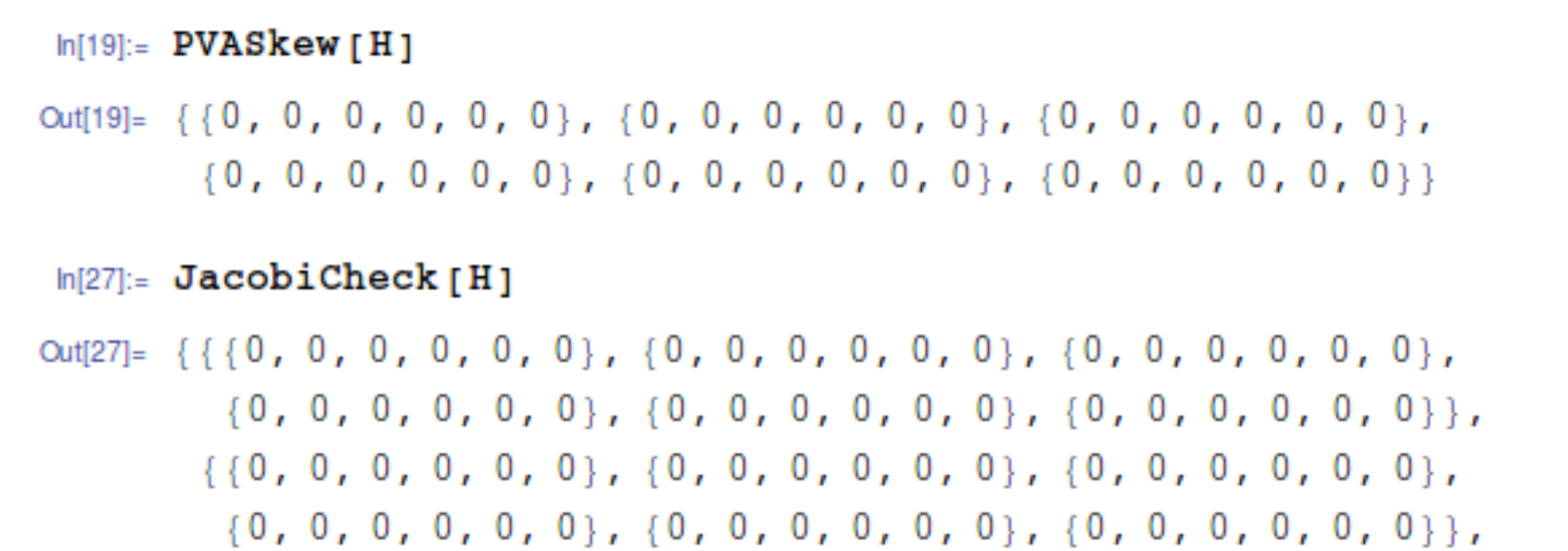}
\end{flushleft}
We use our program to check identity \eqref{20140228:eq4}. The Virasoro element $L$ defined in Proposition
\ref{20160203:prop1}(c) is computed with the command \cmd{GetVirasoro[]}, whose argument is the nilpotent element $f$.
\begin{flushleft}
\hspace{1mm}\includegraphics[scale=0.80]{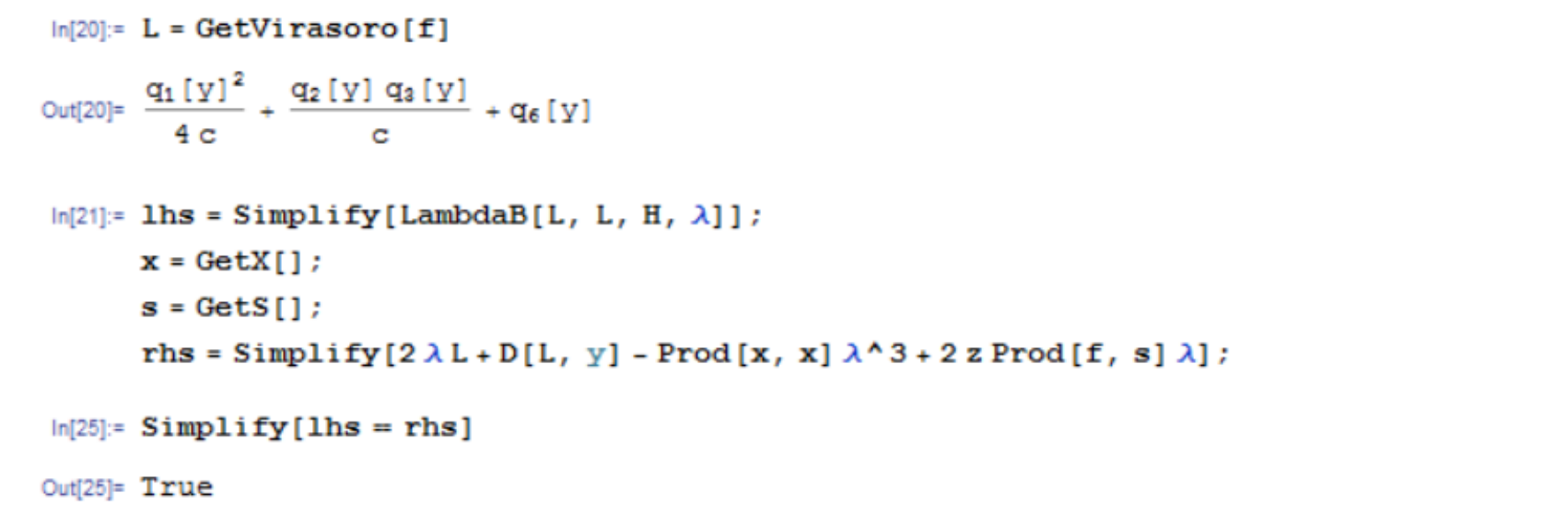}
\end{flushleft}

Finally, we can use our program to compute the first few equations of the corresponding
generalized Drinfeld-Sokolov hierarchies. We define \cmd{g0} and \cmd{g1} according to \cite[Section 6.2]{DSKV14} and we compute the Hamiltonian equation \eqref{hameq}.
We get
\begin{flushleft}
\hspace{1mm}\includegraphics[scale=0.70]{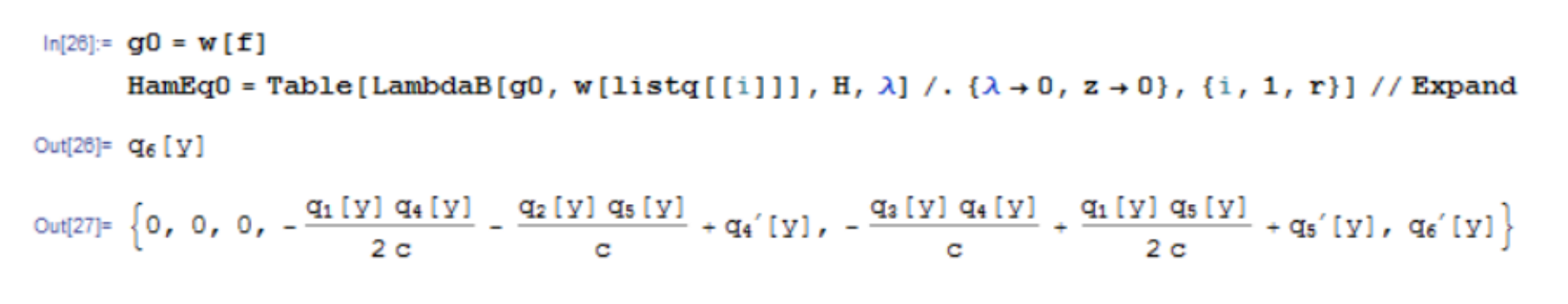}
\end{flushleft}
and
\begin{flushleft}
\hspace{1mm}\includegraphics[scale=0.70]{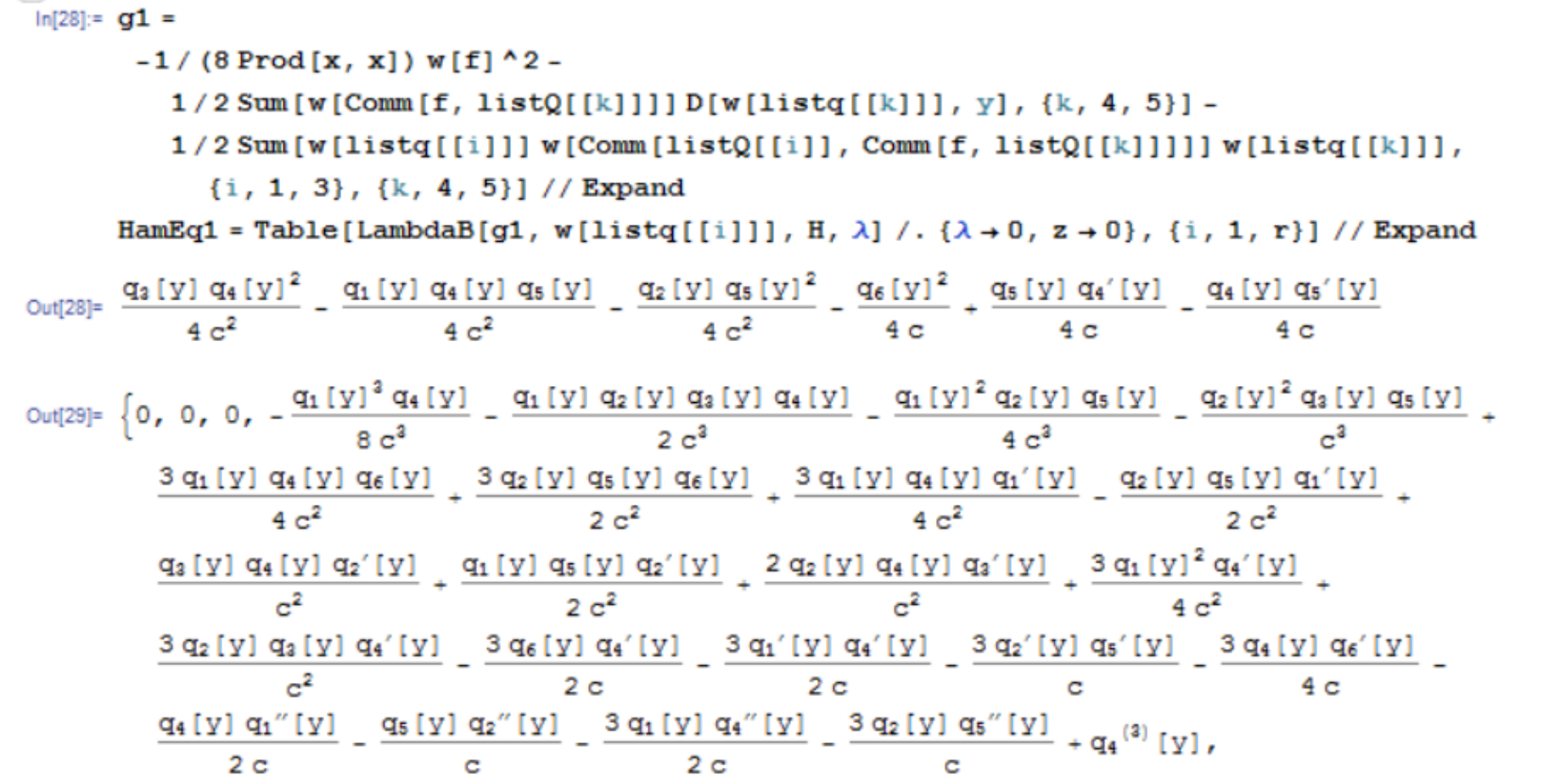}
\end{flushleft}
\begin{flushleft}
\hspace{1mm}\includegraphics[scale=0.70]{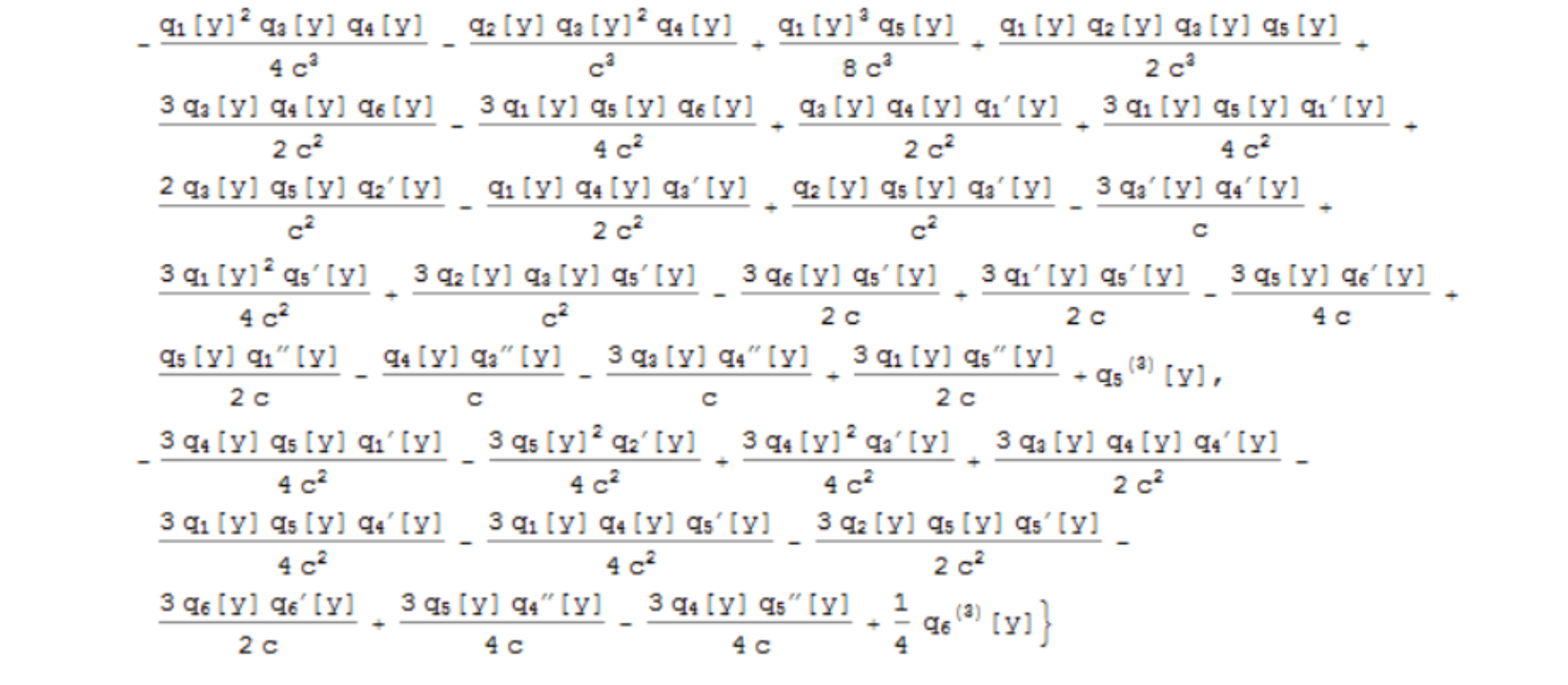}
\end{flushleft}
The above equations agree with equations (6.19) and (6.20) in \cite{DSKV14}.
After a Dirac reduction (since the generators $w(a)$, where $a\in\mf g_0^f$, do not evolve in time), we get simpler equations
\begin{flushleft}
\hspace{1mm}\includegraphics[scale=0.75]{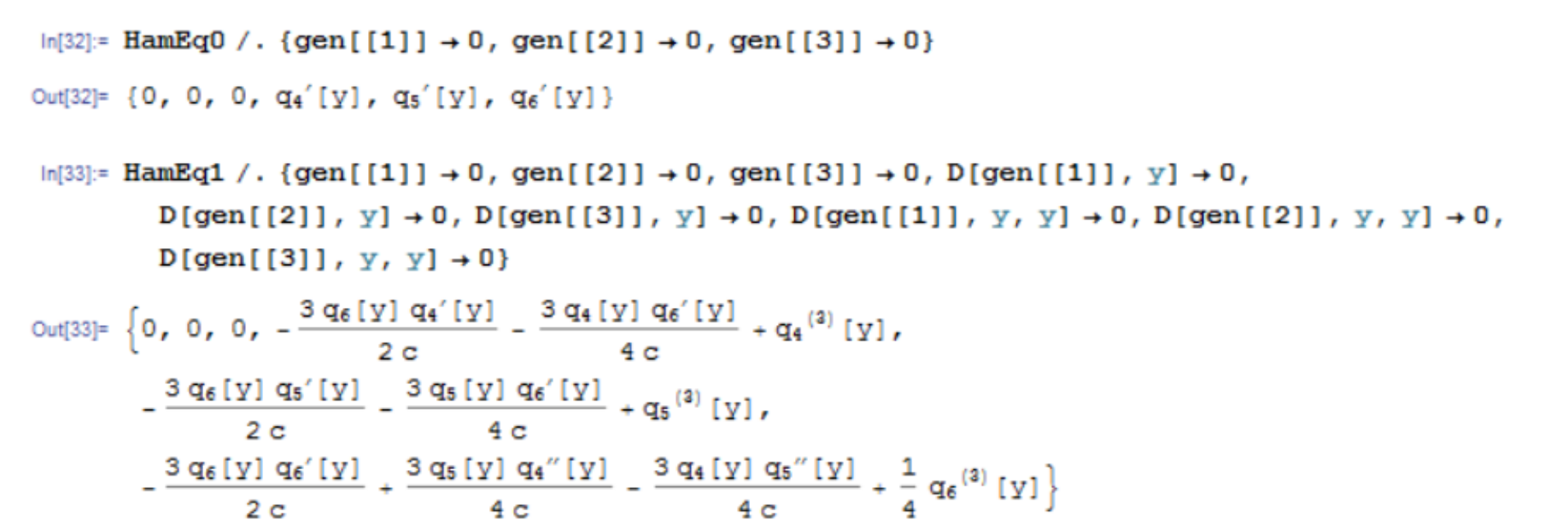}
\end{flushleft}
The above results agree with equations (6.21) and (6.22) in \cite{DSKV14}.
The latter is a higher symmetry of the Yajima-Oikawa equation \cite{YO76} (see also \cite{DSKV14-Err}).

\subsection{Classical affine $\mc W$-algebras associated to simple Lie algebras of rank two and
principal nilpotent element}

In this section we provide explicit formulas for the $\lambda$-brackets among generators
for classical affine $\mc W$-algebras $\mc W(\mf g,f)$, where $\mf g=A_2,B_2$ and $G_2$ and $f$ is a principal nilpotent. In this case, $\dim\mf g^f=\rank\mf g=2$. Hence, by Theorem
\ref{thm:structure-W}, as a differential algebra we have that $\mc W(\mf g,f)=\mb C[w_1^{(n)},w_2^{(n)}\mid n\in\mb Z_+]$, where 
$w_1=w(q_1)$, $w_2=w(q_2)$ and $\{q_1,q_2\}$ is a basis of $\mf g^f$ as in Section \ref{slod.4}.

\subsubsection{$\mf g=\mf{sl}_3$}
The computations can be found in the file \texttt{A\_2\_principal.nb}. The result is
\begin{align}
\begin{split}\label{eq:A_2}
\{{w_1}_\lambda w_1\}_{z}&=(2\lambda+\partial)w_1-2 c\lambda^3
\,,
\\
\{{w_1}_\lambda w_2\}_{z}&=
(3\lambda+\partial)w_2 +3 c z\lambda
\,,
\\
\{{w_2}_\lambda w_2\}_{z}&=
(2\lambda+\partial)\left(\frac1{3c}w_1^2-\frac{1}{16}w_1''\right)
-(2\lambda+\partial)^3\frac{5}{2^43}w_1+\frac{c}{6}\lambda^3
\,.
\end{split}
\end{align}
Note that after rescaling $c\mapsto-\frac C2$ and setting $L=w_1$, $W=8\sqrt{-6C}w_2$,
equation \eqref{eq:A_2} agrees
with the results in \cite{DSKW10} (only the PVA structure corresponding to $z=0$ is considered
there).

\subsubsection{$\mf g=\mf o_5$}
The computations can be found in the file \texttt{B\_2\_principal.nb}. The result is
\begin{align}
\begin{split}\label{eq:B_2}
&\{{w_1}_\lambda w_1\}_{z}=(2\lambda+\partial)w_1-10 c\lambda^3
\,,
\\
&\{{w_1}_\lambda w_2\}_{z}=
(4\lambda+\partial)w_2 +8 c z\lambda
\,,
\\
&\{{w_2}_\lambda w_2\}_{z}=
(2\lambda+\partial)\left(
\frac{2^23^2}{5^4c^2}w_1^3+\frac{7}{5^2c}w_1w_2-\frac{1}{2^25^3c}(w_1')^2
-\frac{29}{2\cdot5^3c}w_1w_1''\right.
\\
&\left.-\frac1{2^25}w_2''+\frac{3}{2^35^2}w_1^{(4)}
\right)
+(2\lambda+\partial)^3
\left(-\frac{7^2}{2^25^3c}w_1^2-\frac{3}{2^25}w_2+\frac{7}{2^25^2}w_1''
\right)
\\
&+(2\lambda+\partial)^5\frac{7}{2^35^2}w_1-\frac{2c}{5}\lambda^7
+z\left((2\lambda+\partial)\frac{2\cdot7}{5^2}w_1-\frac{2^23c}{5}\lambda^3
\right)
\,.
\end{split}
\end{align}
Note that after rescaling $c\mapsto-\frac{C}{10}$ and setting $L=w_1$, $W=-40C\sqrt2w_2$,
equation \eqref{eq:B_2} agrees
with the results in \cite{DSKW10}. Since $\mf o_5\cong\mf{sp}_4$, the corresponding
classical affine $\mc W$-algebras are isomorphic. In fact we can perform the same computations
starting from the Lie algebra $\mf{sp}_4$, which can be found in the file \texttt{C\_2\_principal.nb}, and check that
we get the same expression for the $\lambda$-brackets given by equation \eqref{eq:B_2}
after rescaling $c$ by a factor $\frac12$.

\subsubsection{$\mf g=G_2$}
The computations can be found in the file \texttt{G\_2\_principal.nb}. The result is
\begin{align}
\begin{split}\label{eq:G2}
\{{w_1}_\lambda w_1\}_{z}&=
(2\lambda+\partial)w_1-28c\lambda^3
\,,
\\
\{{w_1}_\lambda w_2\}_{z}&=
(6\lambda+\partial)w_1+144cz\lambda
\,,
\\
\{{w_2}_\lambda w_2\}_{z}&=\sum_{i=0}^4(2\lambda+\partial)^{2i+1}P_{2i+1}
-\frac{3c}{7}\lambda^{11}+z\left(\sum_{i=0}^1(2\lambda+\partial)^{2i+1}Q_{2i+1}
-\frac{26c}{7}\lambda^5\right)
\,,
\end{split}
\end{align}
where
\begin{align*}
&P_1=
\frac{3^35^2}{7^6 c^4}w_1^5
-\frac{11\cdot13}{2\cdot7^3c^2}w_1^2w_2
-\frac{3\cdot61}{2^57^4c^3}w_1^2 (w_1')^2
+\frac{5}{2^37^2c}w_1'w_2'
-\frac{3\cdot769}{2\cdot7^5c^3}w_1^3w_1''
\\
&+\frac{3\cdot29}{2^37^2c}w_2w_1''
+\frac{3^211\cdot19}{2^87^4c^2}(w_1')^2w_1''
+\frac{3^223\cdot97}{2^67^4c^2}w_1(w_1'')^2
+\frac{5}{2^37c}w_1w_2''
\\
&
+\frac{3\cdot347}{2^77^4c^2}w_1w_1'w_1'''
-\frac{3^2}{2^87^2c}(w_1''')^2+\frac{3\cdot9551}{2^87^4c^2}w_1^2w_1^{(4)}
-\frac{3^2\cdot607}{2^87^3c}w_1''w_1^{(4)}
\\
&
-\frac{1}{2^47}w_2^{(4)}-\frac{3^25}{2^87^3c}w_1'w_1^{(5)}
-\frac{3\cdot5\cdot6\cdot7}{2^87^3c}w_1w_1^{(6)}+\frac{3^25}{2^{10}7^2}w_1^{(8)}
\,,\\
&P_3=-\frac{3\cdot11\cdot479}{2^87^4c^3}w_1^4+\frac{5\cdot31}{2^37^2c}w_1w_2
+\frac{3\cdot5\cdot11\cdot19}{2^87^4c^2}w_1(w_1')^2
+\frac{3\cdot11\cdot23\cdot89}{2^77^4c^2}w_1^2w_1''
\\
&
-\frac{3^311\cdot49}{2^87^3c}(w_1'')^2-\frac{5}{2^37}w_2''-\frac{3\cdot5\cdot11}{2^77^3c}w_1'w_1'''
-\frac{3\cdot5^211^2}{2^87^3c}w_1w_1^{(4)}+\frac{3\cdot5\cdot11}{2^87^2}w_1^{(6)}
\,,
\\
&P_5=\frac{3\cdot5\cdot11\cdot139}{2^87^4c^2}w_1^3-\frac{13}{2^47}w_2
-\frac{3^311}{2^97^3c}(w_1')^2-\frac{3^311\cdot43}{2^87^3c}w_1w_1''
+\frac{3^411}{2^97^2}w_1^{(4)}
\,,\\
&P_7=-\frac{3^211\cdot31}{2^97^3c}w_1^2+\frac{3^311}{2^87^2}w_1''
\,,
\qquad\qquad\quad\,\,
P_9=\frac{3\cdot5\cdot11}{2^{10}7^2}w_1
\,,\\
&Q_1=-\frac{11\cdot13}{2\cdot7^3c}w_1^2+\frac{3\cdot29}{2^37^2}w_1''
\,,
\qquad\qquad
Q_3=\frac{5\cdot31}{2^37^2}w_1
\,.
\end{align*}
The bi-Poisson structure of the classical $\mc W$-algebra $\mc W(G_2,f)$ associated to the Lie algebra $G_2$
and its principal nilpotent element $f$, and the corresponding Drinfeld-Sokolov hierarchy
was already computed in \cite{CDVO08}.
It can be obtained from the bi-Poisson structure \eqref{eq:G2} by performing the change of variables
$$
u_0=\frac{w_1}{2c}\,,
\qquad
u_1=\frac{w_2}{72c}+\frac{3}{686}\left(\frac{w_1}{2c}\right)^3
-\frac{33}{1568}\left(\frac{w_1'}{2c}\right)^2-\frac{13}{392}\frac{w_1}{2c}\frac{w_1''}{2c}
+\frac{1}{57}\frac{w_1^{(4)}}{2c}
\,,
$$
by choosing $c=4$, and by substituting $z$ with $-72z$. 

\section{List and explanation of commands}

\subsection{List of commands in MasterPVA}\label{sec:4}
In this section we list the commands provided by \textsl{MasterPVA} and \textsl{MasterPVAmulti}. Most of the commands are the same for both the versions of the package, and the syntax working for the $D=1$ case works the same also when using the multidimensional package; on the other hand, it must be modified accordingly when working with a $D>1$ $\lambda$ bracket.

 \cmd{SetN[n\_Integer]} declares the number $N$ of the generators for the PVA. Its default value is $1$.
 
 \smallskip
 
 \cmd{GetN[]} gives the number $N$ of the generators.
 
 \smallskip
 
 \cmd{SetD[d\_Integer]} declares the number $D$ of the derivations (namely, of the independent variables) for the PVA. Its defaut value is $1$. \emph{Available only in \textsl{MasterPVAmulti}}.
 
 \smallskip
 
 \cmd{GetD[]} gives the number $D$ of the derivations. \emph{Available only in \textsl{MasterPVAmulti}}.
 
 \smallskip
 
 \cmd{SetMaxO[n\_Integer]} declares the order of the derivatives of the generators up to which the code computes the $\lambda$ brackets by the Master Formula. Default is 5, quite high for most of the applications.
 
 \smallskip
 
 \cmd{GetMaxO[]} gives the maximum order of the derivatives of the generators taken by the program.
 
 \smallskip
 
 \cmd{SetGenName[newname]} declares the name for the generators. Default is $u$. They will have the form $u(x)$ if $N=1$ or $u_1(x),\ldots,u_N(x)$ for $N>1$.
 
 \smallskip
 
 \cmd{GetGenName[]} gives the name used for the generators.
 
 \smallskip
 
 \cmd{SetVarName[newname]} declares the name for the independent variable(s). Default is $x$.
 
 \smallskip
 
 \cmd{GetVarName[]} gives the name used for the independent variable(s).
 
 \smallskip
 
 \cmd{gen} is the list of generators for the PVA.
 
 \smallskip
 
 \cmd{var} is the list of the independent variables. \emph{Available only in \textsl{MasterPVAmulti}}.
 
 \smallskip
 
 \cmd{SetFormalParameter[newname]} declares the name for the parameter to be used (and recognized by the software) in the definition of the bracket between generators. Default is $\beta$; notice that for $D>1$ the parameter will be a list $(\beta_1,\ldots,\beta_D)$.
 
 \smallskip
 
 \cmd{GetFormalParameter[]} gives the name of the parameter used in the definition of the $\lambda$ bracket.
 
 \smallskip
 
 \cmd{LambdaB[f,g,P,$\lambda$]} computes the $\lambda$ bracket between the two differential polynomials $f$ and $g$, with $P$ the matrix of the brackets between the generators. The result will be a polynomial in the formal indeterminate $\lambda$ (or $(\lambda_1,\ldots,\lambda_D)$ for $D>1$). The Master Formula will take into account the derivatives of the generators up to order \cmd{n=GetMaxO[]}.
 
 \smallskip
 
 \cmd{PVASkew[P]} computes the condition of skewsymmetry for a $\lambda$ bracket
(namely the LHS of \eqref{skewsimgen}) and gives the result in a matrix form.

\smallskip

\cmd{PrintPVASkew[P]} computes the condition of skewsymmetry and gives the result as a table with each equation of the system.

\smallskip

\cmd{JacobiCheck[P]} computes the $\text{LHS}$ of the Jacobi identity \eqref{jacobigen}, and gives the result as a $N\times N\times N$ array. The entries are given as formal polynomials in the (internal) indeterminates $\lambda$ and $\mu$. It is often convenient to clean up the result using the command \cmd{\%\%//.\{MasterPVA`Private`$\lambda$ ->$\lambda$, MasterPVA`Private`$\mu$ ->$\mu$\}}.

\smallskip

\cmd{PrintJacobiCheck[P]} computes the LHS of Jacobi identity \eqref{jacobigen} and gives the result as a table of expressions that must vanish.

\smallskip

\cmd{EvVField[X\_List,f]} applies the evolutionary vector field of characteristic $X^i$, $i=1,\ldots,N$ to the differential polynomial $f$.

\smallskip

\cmd{Integr[f,param\_List]} transforms a polynomial in the indeterminates \cmd{param}$=\{\lambda,\mu,\ldots,\psi,\omega\}$ in a polynomial in $\{\lambda,\ldots,\psi\}$ substituting $\omega$ with $-\lambda-\mu-\cdots-\psi-\dev$, where $\partial$ acts on the coefficients. For $D>1$ case, each of the parameter must be replaced by a list of $D$ entries. This auxiliary function is convenient in the study of the skewsymmetry, since $\{{u_i}_{-\lambda-\dev}u_j\}$ can be obtained by \cmd{Integr[LambdaB[gen[[i]],gen[[j]],P,$\mu$],\{$\lambda$,$\mu$\}]} or for the study of the PVA cohomology (see \cite{DSK13}).

\subsection{List of commands in WAlg}\label{sec:6}
In this section we list the commands provided by \textsl{WAlg}. We discuss separately the commands that constitute the main core of the program and the ones that can have broader applications, for instance to prepare the input the program needs.

Please note that the symbols \verb=q=, \verb=y=, \verb=z=, \verb=\[ScriptS]= (i.~e.~\emph{s}) and \verb=\[Beta]= (i.~e.~$\beta$) are used by the program, hence they should not be used as variable names in your program.

\subsubsection{Principal commands of the program}
\cmd{InitializeWAlg[name\_String,n\_Integer]} is the first command that the program must receive after loading the package. It sets the simple Lie algebra $\mathfrak{g}$ underlying the classical affine $\mathcal{W}$-algebra. If, for instance, one would like to start from $A_6$, the command should be \cmd{InitializeWAlg["A",6]}.

\smallskip

\cmd{SetNil[a\_List]} sets the nilpotent element $f\in\mathfrak{g}$ in order to
construct $\mc W(\mf g,f)$.

\smallskip

\cmd{GetNil[]} gives the nilpotent element $f$ used in the definition of the classical affine $\mathcal{W}$-algebra.

\smallskip

\cmd{GetDim[]} gives the dimension of the matrices used for the explicit representation of $\mathfrak{g}$.

\smallskip

\cmd{SetS[s\_List]} sets the element $s\in\mf g$ used in the definition of the affine PVA, as in \eqref{lambda}. If the command is given without argument, it authomatically chooses a generic element of $\mathfrak{g}_d$. Notice that the command must be called before computing the $\lambda$-brackets between the generators of the classical affine
$\mathcal{W}$-algebra.

\smallskip

\cmd{GetS[]} gives the element $s$ used in \eqref{lambda} after it has been set.

\smallskip

\cmd{ComputeWAlg[nil\_List]} computes a basis for $\mathfrak{g}^f$ made by $\ad x$-eigenvectors,
where $h=2x$ is the diagonal element of the $\mathfrak{sl}_2$-triple containing $f=\cmd{nil}$, as well as
the dual basis (with respect to the trace form) of $\mf g^e$ and the corresponding $\ad x$-eigenvalues (with multipliciities).

All these outputs can be displayed by using the next three commands:

\smallskip

\cmd{GetWBasis[]} gives the list of elements of the aforementioned basis for $\mathfrak{g}^f$;

\smallskip

\cmd{GetWBasisDual[]} gives the list of elements of the dual basis of $\mf g^e$;

\smallskip

\cmd{GetWEigen[]} gives the list of $\ad x$-eigenvalues.

\smallskip

\cmd{GetX[]} gives the element $x=h/2$, where $h$ is the diagonal element of the $\mathfrak{sl}_2$-triple associated to $f$; it can be used only after executing the command \cmd{ComputeWAlg[]}.

\smallskip

\cmd{w[a\_List]} given an element $a\in\mf g$, it applies the projection $\pi_{\mf g^f}$ to it and then the map $w$ defined in Theorem \ref{thm:structure-W}.

\smallskip

\cmd{GenerateH[par\_]} must be run after \cmd{ComputeWAlg[]} and \cmd{SetS[]}. It computes the
Poisson structure $H$ defined by equation \eqref{ham_W} using Theorem \ref{20140304:thm}.
It uses \cmd{par} as the formal indeterminate (the default is $\beta$).

\smallskip

\cmd{LoadTableIndices[filename\_String]} chooses a file different from the default (\verb=listK_6.txt=) as the source of the indices used in the formula \eqref{20140304:eq4}. It is necessary to use it (after generating the suitable file) when
$d>6$, see Section \ref{sec:5.3}.

\smallskip

\cmd{GenerateTableIndices[n\_Integer]} computes a custom list of indices going up to \verb=n=, and saves it in the file \verb=listK_n.txt= for further usage. Notice that the computation is extremely time-consuming, see Section \ref{sec:5.3}.

\smallskip

\cmd{GetVirasoro[nil\_List]} provides the Virasoro element of Proposition \eqref{20160203:prop1}(c)
with $f=\cmd{nil}$.

\subsubsection{Other useful commands}
\cmd{Comm[a\_,b\_]} computes the commutator between the two matrices \verb=a= and \verb=b=.

\smallskip

\cmd{Prod[a\_,b\_]} computes the value of the symmetric invariant bilinear form $\tr(\mathtt{a}\mathtt{b})$.

\smallskip

\cmd{Proj[a\_List]} applies the map $\pi_{\mf g^f}$ to an element of $a\in\mf g$. It must be run after \cmd{ComputeWAlg}.

\smallskip

\cmd{M[i\_Integer,j\_Integer]} gives the elementary matrix (of dimensions \cmd{GetDim[]}) with 1 in the position $(\mathtt{i},\mathtt{j})$.

\smallskip

\cmd{CheckAlg[a\_List]} checks whether the matrix \verb=a= belongs to the Lie algebra declared in \verb=InitializeWAlg[]=.

\smallskip

\cmd{Sigma[a\_List]} computes $\sigma(\mathtt{a})$ according to the definition given in Section \ref{sec:WAlg}.

\smallskip

\cmd{SetDispPar[s\_]}] sets a dispersive parameter (default is 1, hence making it invisible) in formula
\eqref{20140304:eq4}, useful if we want to compute the dispersionless limit of this formula.

\smallskip

\cmd{GetDispPar[]} gives the aforementioned dispersive parameter.

\subsection{Generation of the indices}\label{sec:5.3}
For $a\in\mf g^f_{-l}$ and $b\in\mf g^f_{-m}$, formula \eqref{20140304:eq4} involves a long summation over the indices $(\vec{j},\vec{n})\in J_{-\vec{k}}$, where $\vec{k}=(k_1,\ldots,k_t)$, $-l+1\leq k_1\prec\cdots\prec k_t\leq m$.
For each $t\geq1$, the list of the indices $k_1,\ldots,k_t$ is finite, and moreover given $l$ and $m$ the maximum value for $t$ is the first integer $\bar t$ such that $\bar t\geq l+m$.

The generation of the indices $\vec{k}$ is a long process, and it dramatically slows down the execution time of the command \cmd{GenerateH[]}. To prevent this issue, a precompiled list of indices is distributed together with the package, in the file \cmd{listK\_6.txt}. It contains the default data for the computation of formula \eqref{20140304:eq4}, and it works for
all classical affine $\mc W$-algebras with $d\leq6$ (this is sufficient, for example, to compute the classical affine $\mc W$-algebras associated to all nilpotent orbits of $\mf o_8$).
If one wants to work with Lie algebras with $d>6$, then it is necessary to generate a bigger table of indices, that may be computed before starting the computation of the Poisson structure $H$, and not necessarily in an interactive session. In case the user does not notice that a bigger set of indices would be needed, the command \cmd{GenerateH[]} will produce a long list of error messages.

The Mathematica kernel, without the user interface, can be usually run in a shell with the command \cmd{math}. After loading the package, one generates the table of indices with the command
\begin{center}\cmd{GenerateTableIndices[d]}.\end{center}
The command can take up to several hours to be completed, and generates a file \cmd{listK\_d.txt} saved in the active folder.
To use a previously generated table of indices, the command \cmd{LoadTableIndices[filename]} must be run before \cmd{GenerateH[]}. The file will be looked for in the active folder, unless the full path is specified.
\begin{flushleft}
\hspace{1mm}\includegraphics[scale=0.55]{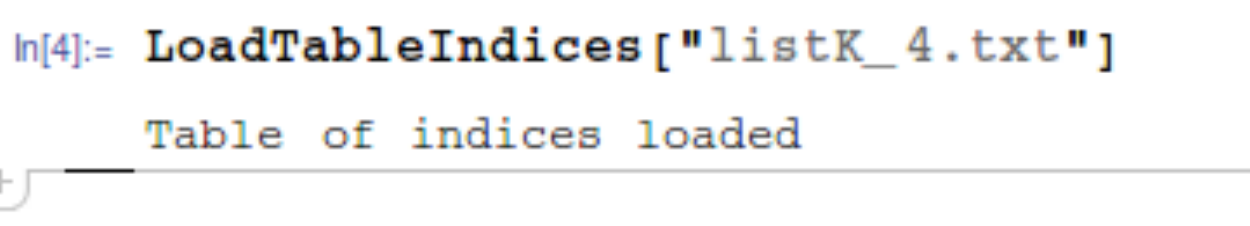}
\end{flushleft}


\end{document}